\documentclass[preprint,12pt, a4paper]{elsarticle}

\usepackage{amssymb}
\usepackage{hyperref}
\usepackage{float}
\usepackage{booktabs}
\usepackage{colortbl}
\usepackage{xcolor}
\usepackage{color}
\usepackage{fancyvrb}

\DefineVerbatimEnvironment{Highlighting}{Verbatim}{commandchars=\\\{\}}
\usepackage{framed}
\definecolor{shadecolor}{RGB}{248,248,248}
\newenvironment{Shaded}{\begin{snugshade}}{\end{snugshade}}

\newcommand{\AttributeTok}[1]{\textcolor[rgb]{0.77,0.63,0.00}{#1}}

\newcommand{\CommentTok}[1]{\textcolor[rgb]{0.56,0.35,0.01}{\textit{#1}}}

\newcommand{\ConstantTok}[1]{\textcolor[rgb]{0.00,0.00,0.00}{#1}}

\newcommand{\DecValTok}[1]{\textcolor[rgb]{0.00,0.00,0.81}{#1}}

\newcommand{\FunctionTok}[1]{\textcolor[rgb]{0.00,0.00,0.00}{#1}}

\newcommand{\NormalTok}[1]{#1}

\newcommand{\OtherTok}[1]{\textcolor[rgb]{0.56,0.35,0.01}{#1}}

\newcommand{\SpecialCharTok}[1]{\textcolor[rgb]{0.00,0.00,0.00}{#1}}

\newcommand{\StringTok}[1]{\textcolor[rgb]{0.31,0.60,0.02}{#1}}

\providecommand{\tightlist}{%
	\setlength{\itemsep}{0pt}\setlength{\parskip}{0pt}}

\journal{arXiv}

\begin{document}

\begin{frontmatter}



\title{cleanTS: Automated (AutoML) Tool to Clean Univariate Time Series at Microscales}


\author[a1]{Mayur Kishor Shende}
\author[a2]{Andrés E. Feijóo-Lorenzo}
\author[a3]{Neeraj Dhanraj Bokde}
\cortext[mycorrespondingauthor]{Corresponding author}
\ead{neerajdhanraj@cae.au.dk}

\address[a1]{Goverment College of Engineering, Nagpur, India}
\address[a2]{Departamento de Enxeñería Eléctrica, Universidade de Vigo, EEI, Campus de Lagoas-Marcosende,  36310 Vigo, Spain}
\address[a3]{Department of Civil and Architectural Engineering, Aarhus University, Denmark}

\begin{abstract}
Data cleaning is one of the most important tasks in data analysis processes. One of the perennial challenges in data analytics is the detection and handling of non valid data. Failing to do so can result in inaccurate analytics and unreliable decisions. The process of properly cleaning such data takes much time. Errors are prevalent in time series data. It is usually found that real world data is unclean and requires some pre-processing. The analysis of large amounts of data is difficult. This paper is intended to provide an easy to use and reliable system which automates the cleaning process of univariate time series data. Automating the process greatly reduces the time required. Visualizing a large amount of data at once is not very effective. To tackle this issue, an R package \emph{cleanTS} is proposed. The proposed system provides a way to analyze data on different scales and resolutions.  Also, it provides users with tools and a benchmark system for comparing various techniques used in data cleaning.

\end{abstract}


\end{frontmatter}

%\linenumbers

\noindent
%\textbf{Main text}\\
% Maximum 3 pages (excluding metadata, tables, figures, references)\\
\section{Introduction}\label{introduction}

Time series data is defined as a sequence of observations taken at successive intervals of time. In an equally spaced time series, the time interval between any two observations is the same. If in a time series only a single variable is varying over time, i.e., only a single type of observation is recorded,
such time series are said to be univariate. It contains the sequence of a single observation, \emph{p\textsubscript{1}, p\textsubscript{2}, 	p\textsubscript{3}, \ldots, p\textsubscript{n}}, recorded at successive points in time, \emph{t\textsubscript{1}, t\textsubscript{2}, t\textsubscript{3}, \ldots, t\textsubscript{n}}. It is usually considered that univariate time series is a single vector of observations, but the time/timestamps can be considered as an implicit variable in the data.

Time series are widely used in many fields \citep{box2015time, brockwell2016introduction, hamiltontime} such as meteorology and hydrology \citep{bokde2019review, gupta2018hybrid}, signal processing, industrial manufacturing, biology \citep{9}, social science \citep{11}, climate observation \citep{13}, pattern recognition, weather forecasting, earthquake prediction, electricity spot price forecasting \citep{cuaresma2004forecasting, bokde2020graphical} and so on. Taylor \citep{10} shows the use of time series in finance, by modeling and forecasting financial time series. Roy et al. \citep{12} use time series in the field of power systems and wind energy. Bokde et al. \citep{bokde2017pattern} explore the suitability of applying pattern similarity-based algorithms to forecast wind speed time series. Besides, various models for short-term wind speed forecasting and power modeling were examined in \citep{bokde2020hybridization}. Chatterjee et al. \citep{chatterjee2020statistical} use univariate time series analysis on COVID-19 datasets for understanding its spread. In many industrial applications, sensors are used to continually record observations over time uninterruptedly \citep{wang2019time}.

\begin{figure}[htbp]
	\centering
	\includegraphics[width=\linewidth]{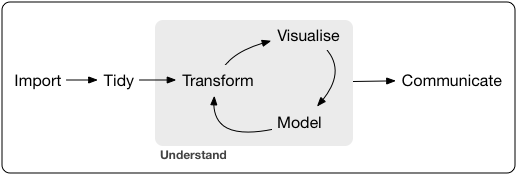}
	\caption{Process of data analysis.}
	\label{fig:figDataScience}
\end{figure}

Data analysis is the process of cleansing, transforming, and modeling data. The goal of data analysis is to derive meaningful and useful information from data. Fig.~\ref{fig:figDataScience} shows the process of data analysis. The first step consists of gathering, importing and cleaning or tidying the data. Then the data is transformed and modeled to get some useful results \citep{r4ds}. Data analysis is used in almost every field of research. It is especially important in business intelligence and analytics. Business intelligence and analytics are data-driven approaches along with processes and tools for extracting information from data \citep{chen2012business, davenport2007competing, lim2013business}. It helps businesses in making well-informed and efficient decisions \citep{davenport2007competing, chaudhuri2011overview}. Data analytics offers a way of analyzing and extracting knowledge and useful insights from the data \citep{davenport2007competing, watson2007current}. Ayankoya et al. \citep{ayankoya2014intrinsic} explain the growing importance of data and data analysis, and the relation between data science, big data, and business analytics. Apart from business intelligence, data analysis is also used in various fields such as risk detection and management, healthcare \citep{lo2016mobile, koh2011data}, security \citep{cardenas2013big}, transportation, and many other.

%\subsection{Data Cleaning}\label{ch1datacleaning}

\begin{figure}
	\centering
	\includegraphics[width=\linewidth]{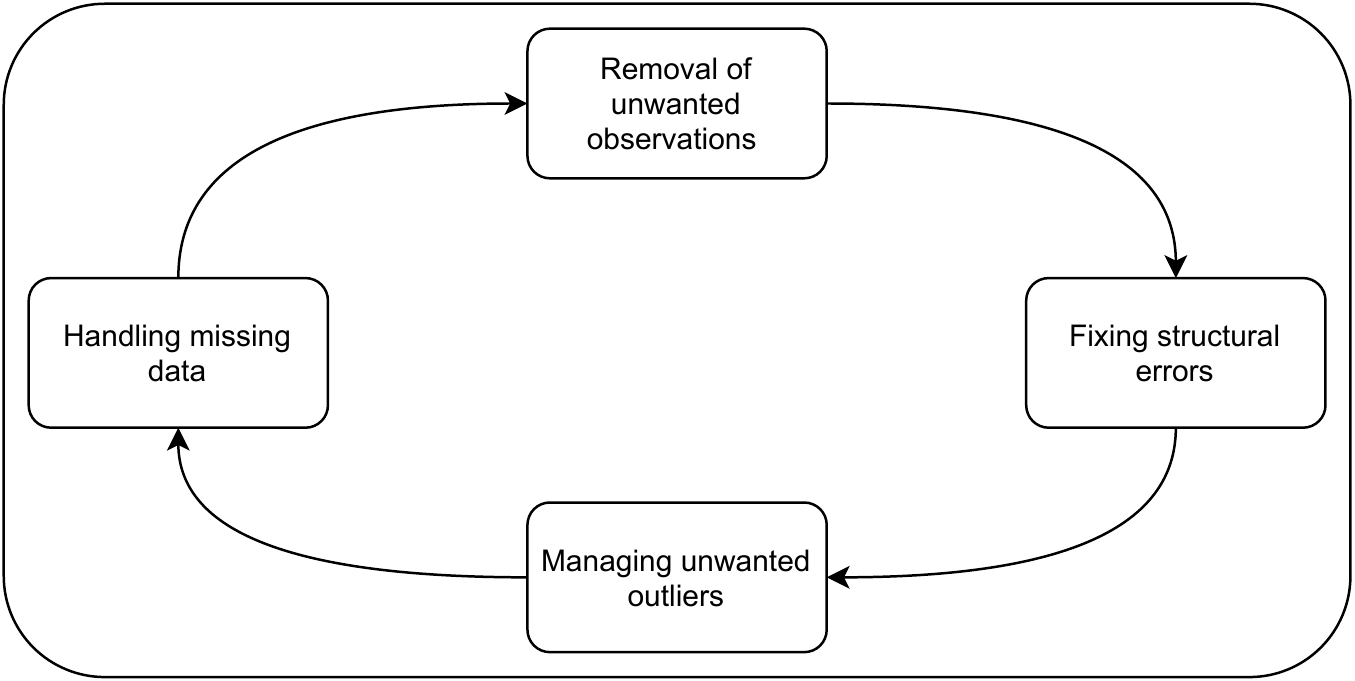}
	\caption{Data cleaning workflow.}
	\label{fig:figDataCleaning}
\end{figure}

Data cleaning is the process of preparing data for analysis by removing or modifying incorrect, incomplete, irrelevant, duplicated, or improperly formatted data. This data is usually not necessary or helpful.%
Fig.~\ref{fig:figDataCleaning} summarizes the process of data cleaning.
Real world data is frequently dirty \citep{jeffery2006declarative} and may contain imprecise values. The same comment is valid for the case with the financial fields \citep{wang2019time}. There may be errors and impurities in the data, which should be filtered out before proceeding to the next steps in data analysis. These impurities can be caused by different factors, varying from faulty equipment, glitches in the systems used for recording observations or errors caused while storing data, to simply human errors. It is possible to reduce errors, but it is impossible to completely avoid them. Dirty time series data may contain impurities such as:

\begin{itemize}
\item	Missing data
\item	Missing timestamps
\item	Outliers
\item	Duplicated observations
\item	Inconsistent data
\item	Problems with data types
\item	Problems with timestamp format, etc.
\end{itemize}

%%%%%%%%%%%%%%%%%%%%%%%%%%%%%%%%%%%%%%%%%%

\section{Motivation}\label{ch1motivation}

Data cleaning is the first step in the data analysis process. The results of all the other steps of the process depend on the results of data cleaning. Therefore, to get a proper analysis of the data it is crucial to clean it. The accuracy of many machine learning data analysis techniques and tools is heavily affected by the data. Many of such algorithms do not work on data containing missing values simply ignored. This may result in the loss of important data. The applications that are built upon unclean data are not reliable, such as pattern mining \citep{morchen2006time, zhang2017time} or classification \citep{xing2012early}. Such data cannot be stored in a database, resulting in loss of data assets. Furthermore, the process of data cleaning is time-consuming and prone to human errors. 

\begin{figure}[htbp]
	\centering 
	\includegraphics[width=0.8\linewidth,]{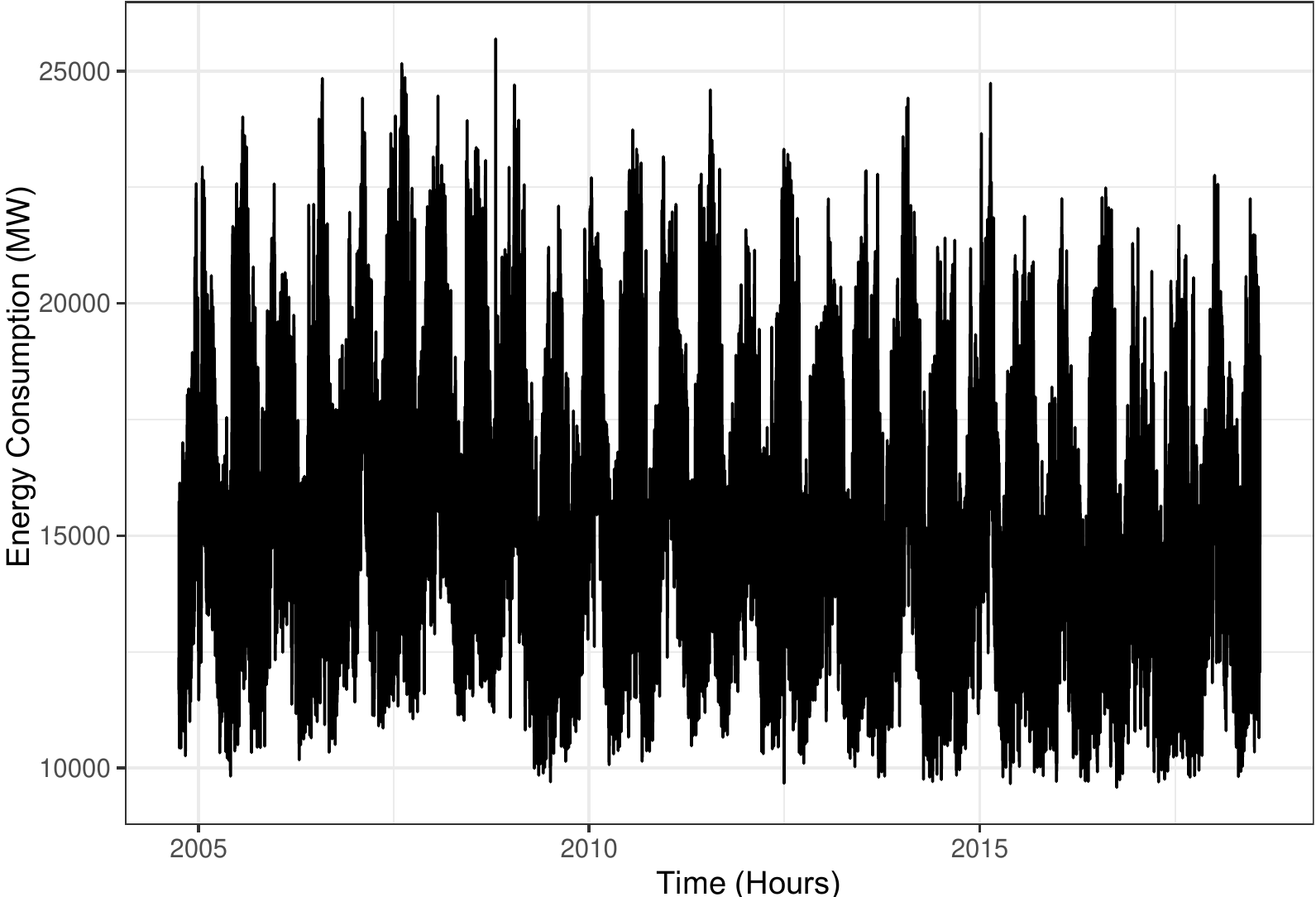} 
	\caption{A sample time-series dataset.}
	\label{fig:figCompleteDataPlot}
\end{figure}

\begin{figure}[htbp]
	\centering 
	\includegraphics[width=0.8\linewidth,]{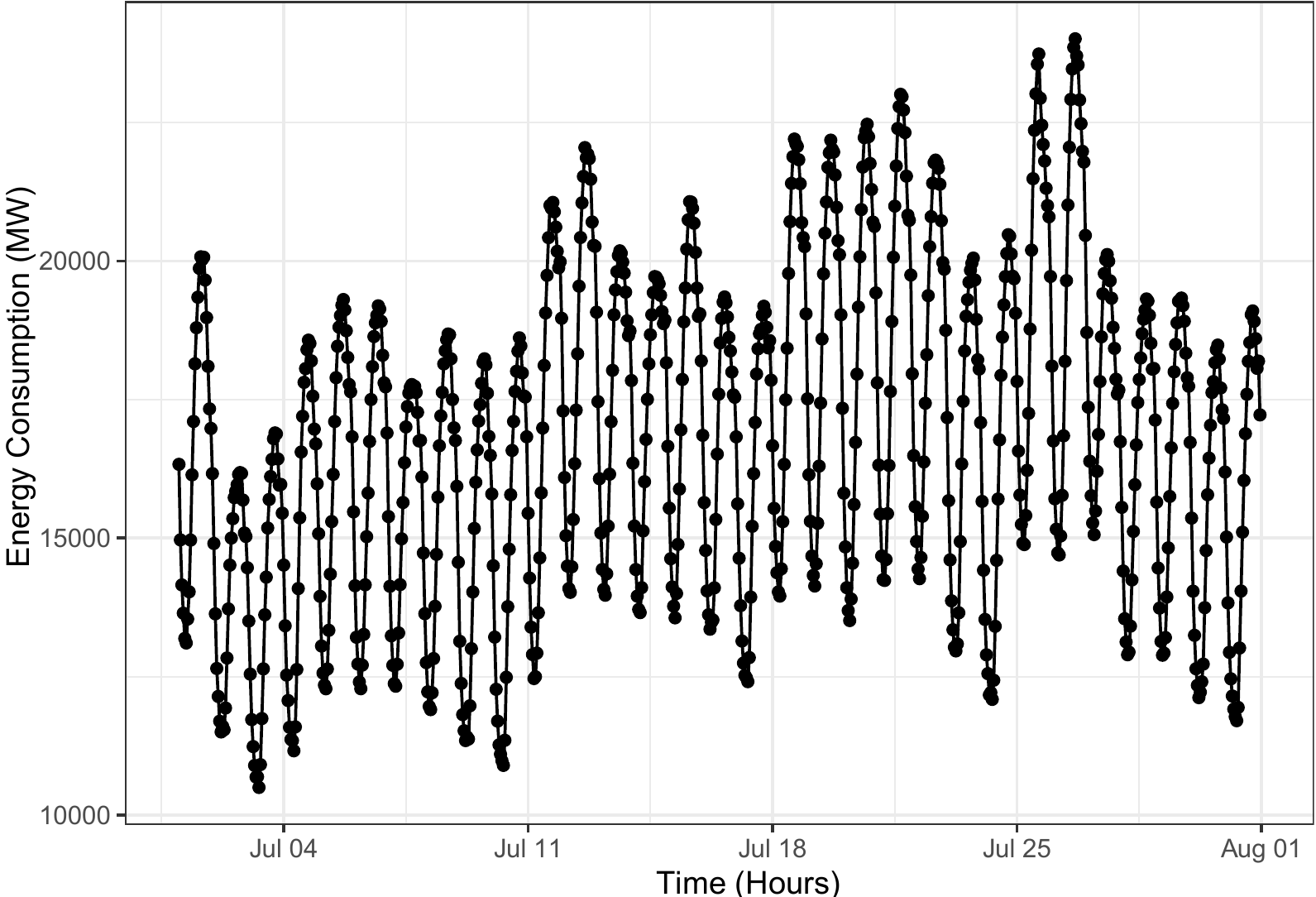}
	\caption{The first month in the dataset shown in Figure \ref{fig:figCompleteDataPlot}.}
	\label{fig:figFrame10}
\end{figure}

The previous sections of this study established the importance of data analysis and time series data cleaning. Therefore, data cleaning should be given great importance when performing data analysis. The data used is growing day by day. There are various tools available for the analysis of big data. Data cleaning and data visualization for such a large amount of data are particularly more challenging. Fig.~\ref{fig:figCompleteDataPlot} shows a time series containing 1,21,273 observations taken from Kaggle (\url{https://www.kaggle.com/robikscube/hourly-energy-consumption}). Since the number of
observations is so high, the patterns in the plot are not visually clear. A subset of this plot is shown in Fig.~\ref{fig:figFrame10}, containing the data for a single month. Viewing the data in the weekly resolution makes it visually clear and more informative. This ensures the importance of analyzing the dataset at microscales. This is a part of the motivation in the visualization strategy for developing the proposed package, called \emph{cleanTS} \citep{cleants}.

%%%%%%%%%%%%%%%%%%%%%%%%%%%%%%%%%%%%%%%%%%
\section{Literature Review} \label{literature}

This section provides a review of state-of-the-art research contributions regarding the data cleaning process. Various tools available for the data cleaning process are discussed in this section. It also explains the concept of missing values, the importance of missing value imputation and several tools and algorithms that have been proposed in the literature and implemented for imputation of missing data.

It is a well established fact that dirty time series data can lead to unreliable and useless analytics, and in fact it has been previously commented. Therefore, data cleaning is the foremost task in the process of data analysis. The different problems that can arise in time series data cleaning are discussed in \citep{wang2019time}. The amount of data and error rate during data collection is high. This is because sensors used to collect data are not always accurate. For example, in a steel mill, the surface temperature of the continuous casting slab cannot be accurately measured or may cause distortion due to the power of the sensor itself. Internet of Things (IoT) data is a common source of time series data. Karkouch et al. \citep{16} explain the details of various IoT data errors generated by various complex environments. Most of the widely used time series cleaning methods utilize the principle of smooth filtering. Such methods may change the original data significantly, and result in the loss of the information contained in the original data. Data cleaning needs to avoid changing the original correct data, a process that should be based on the principle of minimum modification \citep{17, 18, 19}.

\begin{table}

 \caption{\label{tab:cleantools}State-of-the-art data cleaning tools.}
 \centering
 \begin{tabular}{ll>{\raggedright\arraybackslash}p{19em}}
 \toprule
 \textbf{Tool} & \textbf{Method} & \textbf{Description}\\
 \midrule
 \cellcolor{gray!6}{Cleanits} & \cellcolor{gray!6}{Anomaly detection} & \cellcolor{gray!6}{Detects and repair the industrial time-series data. It considers the characteristics of the industrial time-series data and domain-specific knowledge for cleaning.}\\
 EDCleaner & Based on statistics & It works with data related to a social network. The detection and cleaning are performed through the characteristics of statistical data fields.\\
 \cellcolor{gray!6}{TsOutlier} & \cellcolor{gray!6}{Anomaly detection} & \cellcolor{gray!6}{Uses multiple algorithms to detect anomalies in time-series data, and supports both batch and streaming processing.}\\
 ASPA & Smoothing based & Automatically smooths streaming time series by adaptively optimizing the trade-off between noise reduction and trend retention.\\
 \cellcolor{gray!6}{PACAS} & \cellcolor{gray!6}{Based on statistics} & \cellcolor{gray!6}{Design a framework for data cleaning between service providers and customers.}\\
 PIClean & Based on statistics & Produces probabilistic errors and probabilistic fixes which help in implicitly discovering and using relationships between data columns for cleaning.\\
 \cellcolor{gray!6}{HoloClean} & \cellcolor{gray!6}{Based on statistics} & \cellcolor{gray!6}{Learn the probability model and select the data cleaning plan based on probability distribution.}\\
 ActiveClean & Based on statistics & Allows for progressive and iterative cleaning in statistical modeling problems.\\
 \cellcolor{gray!6}{MLClean} & \cellcolor{gray!6}{Anomaly detection} & \cellcolor{gray!6}{Combines data cleaning with machine learning methods.}\\
 \bottomrule
 \end{tabular}
 \end{table}

Data cleaning is an important field for research and there have been various tools and systems proposed for it. Some of them have been listed in \citep{wang2019time}. Ding et al. \citep{20} propose an industrial time series cleaning system, \emph{Cleanits}, which can detect and repair industrial time series. It provides a friendly interface so users can use results and logging visualization over every cleaning process. The algorithms used also take into consideration the characteristics of the industrial time series data and domain-specific knowledge. \emph{EDCleaner} is proposed in \citep{21} and works with data related to the social networks. The detection and cleaning are performed through the characteristics of statistical data fields. \emph{TsOutlier} is a framework for detecting outliers presented in IoT data \citep{22}, that uses multiple algorithms to detect anomalies in time series data, and also supports batch and stream processing. The \emph{ASPA} is a smoothing-based analytics operator that automatically smooths streaming time series by adaptively optimizing the trade-off between noise reduction and trend retention\citep{23}. It violates the minimum modification principle and distorts the data, making it unsuitable for a wide use. \emph{PACAS} \citep{24} is a framework for data cleaning between service providers and customers. \emph{PIClean} \citep{25} is a statistics-based tool for cleaning. It produces probabilistic errors and probabilistic fixes which helps in implicitly discovering and using relationships between data columns for cleaning. \emph{HoloClean} \citep{27} selects the data cleaning plan based on probability distribution. \emph{ActiveClean} \citep{26} allows progressive and iterative cleaning in statistical modeling problems. \emph{MLClean} \citep{28} is an anomaly detection tool, which combines data cleaning with machine learning methods. Table \ref{tab:cleantools} lists all the mentioned tools.

The problem of missing data arises frequently and is very common. A lot of research has been done in the field of imputation. Almost whenever data is recorded, problems regarding missing values occur. There are different reasons for the absence of an observation, such as not measured or lost values or values that have been finally considered not valid \citep{14}. There are three missing data mechanisms, discussed in \citep{14}:

\begin{itemize}
	\item
	\emph{Missing completely at random (MCAR)}:
	In MCAR there is no systematic mechanism on the way the data is missing. The occurrence of missing data points is completely random. This means that in univariate time series data, the probability of the observation to be missed does not depend on the time the observation is recorded.
	\item
	\emph{Missing at random (MAR)}:
	In MAR the probability of missed observation does not depend on the value of the observation itself, but on other variables. As pointed out in \citep{14}, the majority of missing data methods require MAR or MCAR. %since the missing data mechanism is said to be ignorable for them \citep{15} 
	The MAR mechanism allows the imputation algorithms to use correlations with other variables, so better results compared to MCAR can be obtained.
	\item
	\emph{Not missing at random (NMAR)}:
	In NMAR, the data points are not missing at random. The probability of a missed value depends on the value of the observation, and can also be dependent on other variables. NMAR is called non-ignorable because in order to perform the imputation, a special model for why data is missing and what the likely values are, needs to be included.
\end{itemize}

There are various algorithms and packages in the R programming language to deal with missing data. Some of these ar, imputation based on random forests \citep{stekhoven2012missforest}, nearest neighbor observation \citep{crookston2008yaimpute}, predictive mean matching \citep{meinfelderpackage}, maximum likelihood estimation \citep{gross2011mvnmle}, conditional copula specifications \citep{di2014imputation}, expectation-maximization \citep{mtsdi}. \citep{31} and \citep{vim} provide various algorithms and tools for imputation.

An anomaly or an outlier is a recorded observation in a time series data, which is significantly different from other observations. Such an observation deviates too much from other ones. They are also called abnormalities, deviants and discordants \citep{laptev2015generic, aggarwal2015outlier}. Outlier detection is very useful and important in many areas like intrusion detection, credit-card fraud, medical diagnosis, earth science, law enforcement and many more. Anomaly detection is also very important in data analysis. It is possible that a data point which represents an anomaly may be an error while recording the observation, i.e, it is an invalid data point. Such invalid data points are not desirable for data analysis, since they can significantly affect the data analysis results. But it is also possible that the data point is correct. If it is in fact an error, then it is important to remove it from the data before analyzing the data.

%%%%%%%%%%%%%%%%%%%%%%%%%%%%%%%%%%%%%%%%%%
\section{Introduction to R Package \emph{cleanTS}} \label{theRpackage}

\begin{figure}[htbp]
	\centering 
	\includegraphics[width=1\linewidth,]{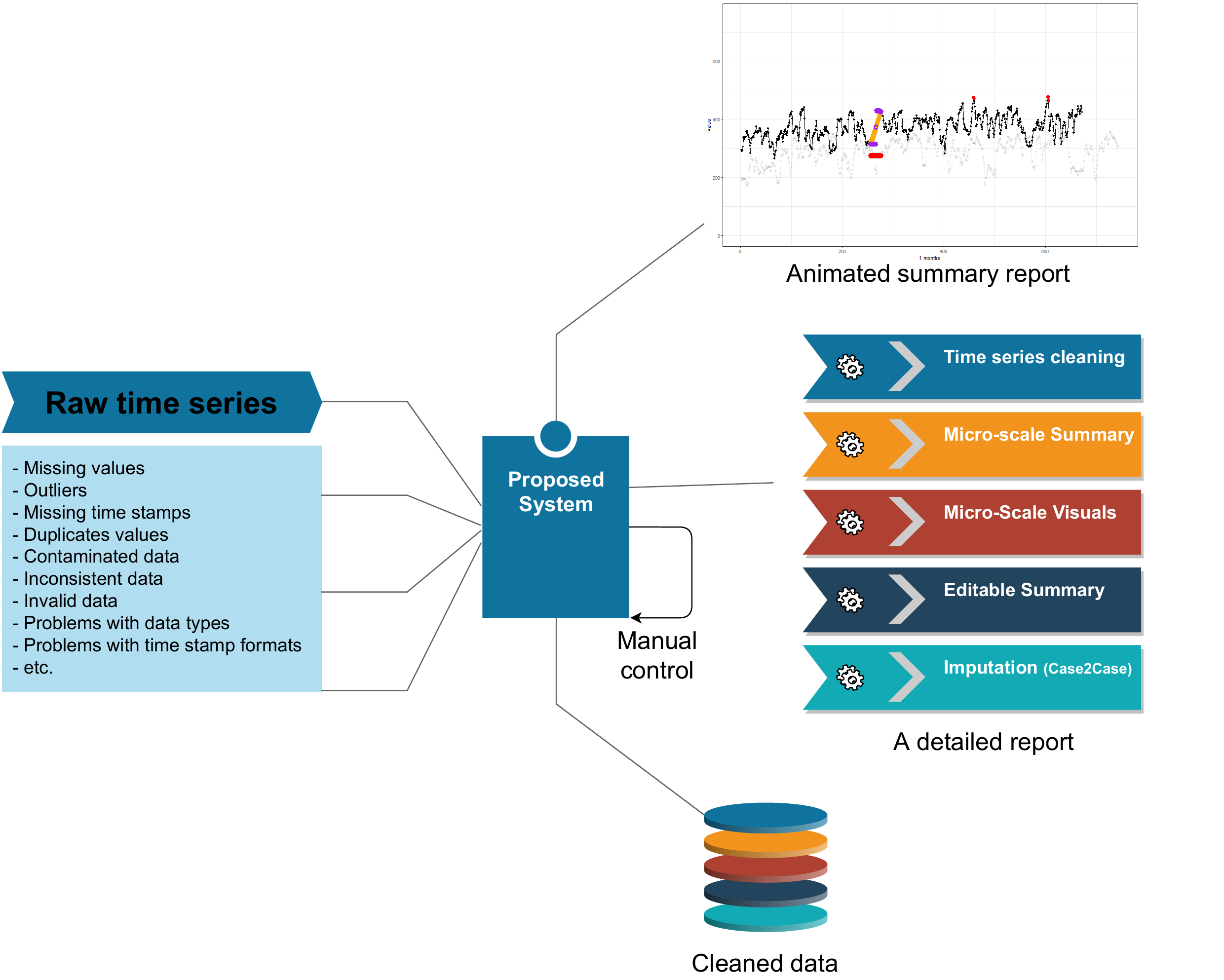} 
	\caption{Worflow of the proposed system.}
	\label{fig:workflow}
\end{figure}

This package focuses on the development of a tool that makes the process of cleaning large datasets simple and time-efficient. It implements reliable and efficient procedures for automating the process of cleaning univariate time series data. The time required for cleaning the data is significantly reduced if the process is automated. The tool provides integration with already developed and deployed tools for missing value imputation. The main problem with visualizing large amounts of data is that the visualizations are not very informative. The tool provides a way of visualizing large time series data in different resolutions. It is intended to be used by researchers from various domains, who want to work on data-science-related projects. Gateways and procedures are also included in the tool, for the researchers who are interested in using the proposed tool for introducing and adding new methodologies and algorithms in the domain. Figure~ \ref{fig:workflow} contains a brief summary of the proposed system. The tool is designed such that it requires minimum user interaction. The ultimate goal is the creation of a handy software tool that deals with all the problems, processes, analysis, and visualization of big data time series, with or without human intervention.

\begin{figure}[htbp]
	\centering 
	\includegraphics[width=1\linewidth,]{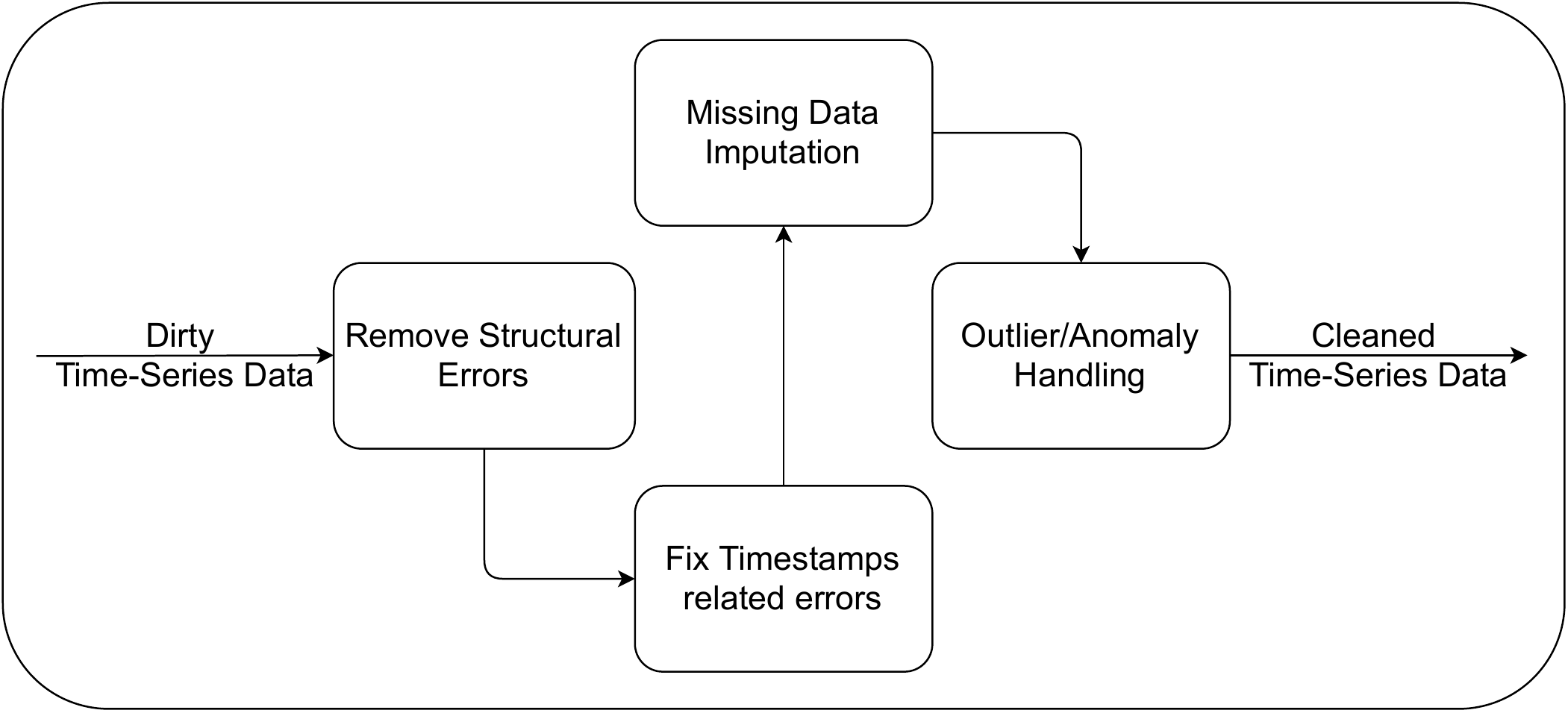} 
	\caption{Flowchart for the proposed system.}
	\label{fig:flowchart}
\end{figure}

Figures \ref{fig:workflow} and \ref{fig:flowchart} show the workflow of the system. The system requires univariate time series data as input. Data cleaning of multivariate time series and non-time series data is out of the scope of the present version of the tool. Section~\ref{introduction} listed the impurities that may be present in the time series data. The system has procedures implemented to handle each of these impurities. After the impurities have been removed or corrected, it generates a detailed report on the entire process, notifying the user of all the changes to make to the original data. This report allows the users to review the changes and revert them make if required. The system also provides a tool for visualizing the data in different resolutions. These procedures generate an animated visualization or an interactive plot, which helps with the micro-scale visualization and analysis of the data.

R is a programming language specifically designed to be used for statistical computing and graphical visualization \citep{citeR}. The graphics tools provided by R are one of the best for data analysis and display either on-screen or on hardcopy. There are many efficient and reliable tools for performing any task related to data science, such as data wrangling, data visualization, machine learning, etc. These packages are updated and maintained regularly. The proposed system is implemented in the R programming language. This section provides details on each function in the package. All the functions available to the users are listed in Table \ref{tab:functionTable}. Each of the data cleaning tasks is divided into internal functions. Several other internal helper functions are not intended to be used directly by the user and hence are not listed here.

In R there are various libraries used for data manipulation, data wrangling, and working with data in general. Two such libraries are the \emph{data.table} package \citep{datatable} and the \emph{tidyverse} family of packages \citep{tidyverse}. The \emph{tidyverse} is a collection of packages for solving data science challenges using R code. Some of the packages in tidyverse includes \emph{dplyr}\citep{dplyr}, \emph{tibble}\citep{tibble}, \emph{ggplot2}\citep{ggplot2} and \emph{tidyr}\citep{tidyr}. They are user-friendly, efficient, and share the same design methodology. Also, the code written with these packages is clean and easily understandable. The \emph{data.table} provides a high-performance version of base R's \texttt{data.frame}. It is useful for tasks such as aggregating, filtering, merging, grouping, and other related tasks. Both of these packages are a lot faster than their base R equivalents. When considering \emph{data.table} and \emph{dplyr}, it can be seen that \emph{data.table} gets faster than \emph{dplyr} as the number of groups and/or rows to group increase \citep{q1}. Since the proposed tool needs to work with a large amount of data, the R package uses the \emph{data.table} backend.

\subsection{Highlights} \label{highlights}
\begin{enumerate}
	\item	\textbf{Automation of data cleaning}: Primarily, the package automates the cleaning and organizing the process of cleaning big (voluminous) time series data. It includes fixing structural errors, timestamp related errors, and handling missing values and anomalies in the data. The process of univariate time series cleaning is discussed in detail in Section~\ref{introduction}. 
	
	\item	\textbf{Integrated with imputation tools}: There are various tools, available for the automation of missing value imputation. These tools are already tested and deployed. The \emph{cleanTS} package makes use of such tools for handling missing value imputation in univariate time series data. One such package used is the \emph{imputeTestbench} package, which provides a benchmarking tool for comparing various methods of imputation. It is also possible to add new imputation methodology and algorithms and compare them to existing once. This integration has enabled the creation of a handy software tool that deal with the pre-processing, analysis and visualization of big data time series with minimum to no human intervention.
	
	\item	\textbf{Graphical user-interface}: The tool is targeted towards researchers working in several domains and willing to work on data science related projects in their respective domains with an interactive tool. It provides a user-friendly and easy to understand GUI (graphical user-interface). This enables the tool to be used by the users with no coding knowledge or experience.

	\item 	\textbf{Micro scale visualization}:The package provides procedures and functions for visualizing the time series data at micro scales. It involves splitting the data according to the provided interval and then creating the visualization for each part of the data. This tool analyzes the time series at the micro-level and assists in cleaning it in an interactive manner with data science principles. 
	
\end{enumerate}	

\subsection{Installation}\label{ch3install}

The \emph{cleanTS} package can be installed from github.

\begin{Shaded}
\begin{Highlighting}[]

    \CommentTok{\# Install from GitHub}
    \FunctionTok{install.packages}\NormalTok{(}\StringTok{"devtools"}\NormalTok{)}
    \NormalTok{devtools}\SpecialCharTok{::}\FunctionTok{install\_github}\NormalTok{(}\StringTok{"Mayur1009/cleanTS"}\NormalTok{)}
    
    \CommentTok{\# Install from CRAN}
    \FunctionTok{install.packages}\NormalTok{(}\StringTok{"cleanTS"}\NormalTok{)}
\end{Highlighting}
\end{Shaded}

%Once the package is submitted to \emph{CRAN}(Comprehensive R Archive Network), it can be directly installed through \emph{CRAN}. \emph{CRAN} is a network of FTP and web servers around the world that store identical, up-to-date, versions of code and documentation for R.
%
%\begin{Shaded}
%\begin{Highlighting}[]
%	\FunctionTok{install.packages}\NormalTok{(}\StringTok{"cleanTS"}\NormalTok{)}
%\end{Highlighting}
%\end{Shaded}

The system on which the package is to be installed needs to have R and RStudio installed on the machine. On a Windows machine, this setup is enough to install the package, but on Linux-based systems like Ubuntu, some extra packages are required. For the installation of the \emph{gganimate} package a Rust compiler is required. This can be installed by using, \texttt{sudo\ apt-get\ install\ cargo} on Debian/Ubuntu, \texttt{yum\ install\ cargo} on Fedora/CentOS, \texttt{brew\ install\ rustc} on MacOS.

\subsection{Functions and Implementation Methodology}\label{ch3functions}

\begin{table}

 \caption{\label{tab:functionTable}Functions in \emph{cleanTS} package.}
 \centering
 \begin{tabular}{>{}l>{\raggedright\arraybackslash}p{22em}}
 \toprule
 \textbf{Functions} & \textbf{Description}\\
 \midrule
 \ttfamily{\cellcolor{gray!6}{cleanTS()}} & \cellcolor{gray!6}{Function for cleaning the input data. It creates and returns a cleanTS object.}\\
 \ttfamily{gen.report()} & Generates a report of the process of data cleaning, from the given `cleanTS` object.\\
 \ttfamily{\cellcolor{gray!6}{animate\_interval()}} & \cellcolor{gray!6}{Create an animated plot from the given cleanTS object and a specified interval.}\\
 \ttfamily{gen.animation()} & Renders the animation using a gganim object returned by animate\_interval().\\
 \ttfamily{\cellcolor{gray!6}{interact\_plot()}} & \cellcolor{gray!6}{Creates an interactive plot from the given cleanTS object and a specified interval.}\\
 \bottomrule
 \end{tabular}
 \end{table}

\begin{enumerate}
	%%%% FUNCTION 1 %%%%
	\item	\underline{\texttt{cleanTS()}} \label{cleanTS}
	
	\begin{Shaded}		
\begin{Highlighting}[]
\FunctionTok{cleanTS}\NormalTok{(data, date\_format, }\AttributeTok{imp\_methods =} \FunctionTok{c}\NormalTok{(}\StringTok{"na\_interpolation"}\NormalTok{, }
	\StringTok{"na\_locf"}\NormalTok{, }\StringTok{"na\_ma"}\NormalTok{, }\StringTok{"na\_kalman"}\NormalTok{), }\AttributeTok{time =} \ConstantTok{NULL}\NormalTok{, }\AttributeTok{value =} \ConstantTok{NULL}\NormalTok{, }
	\AttributeTok{replace\_outliers =}\NormalTok{ T)}
\end{Highlighting}
	\end{Shaded}

	\begin{itemize}
		\item
		\texttt{data}: The input time series data. Can be a \texttt{data.frame}, \texttt{tbl}, or table-like object.
		\item
		\texttt{date\_format}: A character string, the format of the time column in the data.
		\item
		\texttt{imp\_methods}: A vector of strings, the methods of imputation to be used for imputing missing values. The default value specifies four methods, \texttt{na\_interpolation}, \texttt{na\_locf}, \texttt{na\_ma}, \texttt{na\_kalman}.
		\item
		\texttt{time}: Name of the column containing timestamps. If \texttt{NULL} the first column is considered to be the time column.
		\item
		\texttt{value}: Name of the column containing observations. If \texttt{NULL} the second column is considered to be the time column.
		\item
		\texttt{replace\_outliers}: Defaults to \texttt{TRUE}. Specify whether to remove and impute the detected outliers in the time series.
	\end{itemize}
	
	\texttt{cleanTS()} is the entry function to the package. It is a wrapper function that calls all the other internal functions to performs different data cleaning tasks. The first task is to check the input time series data for structural and data type-related errors. Since the functions need univariate time series data, the input data is checked for the number of columns. By default, the first column is considered to be the time column, and the second column to be the observations. Alternatively, if the time and value arguments are given, then those columns are used. The time column is converted to a POSIX object using the \emph{lubridate} package \citep{lubridate}. \emph{Lubridate} allows the format to be specified in a very easy and simplified way. A complete list of all the possible date-time formats is provides in \citep{q2}. The value column is converted to a numeric type. If it contains invalid data, like a string of random characters, which cannot be parsed to numeric, they are replaced with \texttt{NA}. The column names are also changed to \texttt{time} and \texttt{value}. All the data is converted to a \emph{data.table} object. This data is then passed to other functions to check for missing and duplicate timestamps. If there are any missing timestamps found, they are inserted in the data and the corresponding observations are set to \texttt{NA}. If duplicate timestamps are found, then the observation values are checked. If the observations are the same, then only one copy of that observation is kept. But if the observations are different, then it is not possible to find the correct one, so the observation is set to \texttt{NA}.
	
	This data is then passed to a function for finding and handling missing observations. These are represented by \texttt{NA} in the \emph{value} column of the data. The problem of missing data arises frequently and is very common. A lot of research has been done in the field of imputation, which has been discussed in Section~\ref{literature}. The package provides integration with the \emph{imputeTestbench} package \citep{imputeTB, 30}. It provides the function for comparing various methods of imputation. Using these functions the methods given in the \texttt{imp\_methods} argument are compared and selected. The \emph{imputeTestbench} also offers functionality to separately find the best methods for \emph{MCAR} and \emph{MAR} types of missing values. After the best methods are found, imputation is performed using those methods. The user can also pass user-defined functions for comparison. The default functions are provided by the \emph{imputeTS} package \citep{imputeTS}. It provides functions for imputation by linear interpolation, imputation by structural model and Kalman smoothing, imputation by last observation carried forward, imputation by simple moving average, imputation by mean value, and many more. The user-defined function should follow the structure as the default functions. It should take a numeric vector containing missing values as input, and return a numeric vector of the same length without missing values as output.
	
	Once the missing values are handled the data is checked for outliers. The \emph{anomalize} package \citep{anomalize} provides great functions for finding outliers/anomalies in time series data. The \emph{anomalize} package accepts only \texttt{tibble} (class \texttt{tbl\_df}) \citep{tibble} or \texttt{tibbletime} (class \texttt{tbl\_time}) \citep{tibbletime} objects. The \texttt{tibbletime} is an extension of \texttt{tibble} that creates time-aware tibbles by setting a time index. The general workflow for anomaly detection includes the decomposition of the time series data into the seasonal, trend, and remainder components, then applying anomaly detection on the remainder part. This generates the lower and upper limits for the data. Any observation outside these limits is treated as an outlier or anomaly. If the \texttt{replace\_outliers} parameter is set to \texttt{TRUE} in the \texttt{cleanTS()} function, then the outliers are replaced by \texttt{NA} and imputed using the procedure mentioned for imputing missing values. Then it creates a \texttt{cleanTS} object which contains the cleaned data, missing timestamps, duplicate timestamps, imputation methods, MCAR imputation error, MAR imputation error, outliers, and if the outliers are replaced then imputation errors for those imputations are also included. The \texttt{cleanTS} object is returned by the function.\\
	
	%%%%%%%%%%%%%%%%%%%%
	
	%%%% FUNCTION 2 %%%%
	\item	\underline{\texttt{gen.report()}} \label{genrep}
	
	\begin{Shaded}
\begin{Highlighting}[]
\FunctionTok{gen.report}\NormalTok{(obj)}
\end{Highlighting}
	\end{Shaded}
	
	\begin{itemize}
		\tightlist
		\item
		\texttt{obj}: The \emph{cleanTS} object, returned by the \texttt{cleanTS()} function.
	\end{itemize}
	
	The \texttt{cleanTS()} function handles all the data cleaning tasks. It makes a lot of changes to the original data. The \texttt{gen.report()} function shows a report of these changes and gives details about the impurities found in the data.\\
	%%%%%%%%%%%%%%%%%%%%	
	
	%%%% FUNCTION 3 %%%%
	\item	\underline{\texttt{animate\_interval()}} \label{anim}
	
	\begin{Shaded}
\begin{Highlighting}[]
\FunctionTok{animate\_interval}\NormalTok{(obj, interval)}
\end{Highlighting}
	\end{Shaded}
	
	\begin{itemize}
		\tightlist
		\item
		\texttt{obj}: The \emph{cleanTS} object, returned by the \texttt{cleanTS()} function.
		\item
		\texttt{interval}: A string or numeric value, specifying the viewing resolution in the plot.
	\end{itemize}
	
	\texttt{animate\_interval()} creates an animated plot for the given data. First, the data is split according to the \texttt{interval}. If it is a numeric value, the cleaned data is split into dataframes containing \texttt{interval} observations. It can also be a string, like \emph{1 week}, \emph{3 months}, \emph{14 days}, etc. In this case, the data is split according to the \texttt{interval} given. The \emph{gganimate} package \citep{gganimate} is an extension of the \emph{ggplot2} library \citep{ggplot2}, which adds functionality to animate the plot. Here we split the data into states according to the given \texttt{interval} and then use \texttt{transition\_state()} function from \emph{gganimate}. The \texttt{animate\_interval()} function returns a list containing the \texttt{gganim} object used to generate the animation and the number of states in the data. The animation can be generated using the \texttt{gen.animation()} function and saved using the \texttt{anim\_save()} function. The plots in the animation also contain a short summary, containing the statistical information and the number of missing values, outliers, missing timestamps, and duplicate timestamps in the data shown in that frame of animation.\\

	%%%%%%%%%%%%%%%%%%%%
	
	%%%% FUNCTION 4 %%%%
	\item	\underline{\texttt{gen.animation()}} \label{genanim}
	
	\begin{Shaded}
\begin{Highlighting}[]
\FunctionTok{gen.animation}\NormalTok{(anim, }\AttributeTok{nframes =} \DecValTok{2} \SpecialCharTok{*}\NormalTok{ anim}\SpecialCharTok{$}\NormalTok{nstates, }
	\AttributeTok{duration =}\NormalTok{ anim}\SpecialCharTok{$}\NormalTok{nstate, ...)}
\end{Highlighting}
	\end{Shaded}
	
	\begin{itemize}
		\tightlist
		\item
		\texttt{anim}: A list containing a \texttt{gganim} object and number of states (\texttt{numeric}).
		\item
		\texttt{nframes}: The number of frames to render in the animation.
		\item
		\texttt{duration}: The duration of the generated animation.
		\item
		\texttt{...} : Other arguments passed to \texttt{animate()} function in the \emph{gganimate} package.
	\end{itemize}
	
	\texttt{gen.animation()} is a simple wrapper function for the \texttt{animate()} function which is used to render the animation using a \texttt{gganim} object. By default, in the \texttt{animate()} function only 50 states in the data are shown. So, to avoid this \texttt{gen.animation()} defines the default value for the number of frames. Also, the \texttt{duration} argument has a default value equal to the number of states, making the animation slower. More arguments can be passed, which are then passed to \texttt{animate()}, like, height, width, fps, renderer, etc.\\
	
	%%%%%%%%%%%%%%%%%%%%
	
	%%%% FUNCTION 5 %%%%
	\item	\underline{\texttt{interact\_plot()}} \label{intp}
	
	\begin{Shaded}
\begin{Highlighting}[]
\FunctionTok{interact\_plot}\NormalTok{(obj, interval)}
\end{Highlighting}
	\end{Shaded}
	
	\begin{itemize}
		\tightlist
		\item
		\texttt{obj}: The \emph{cleanTS} object, returned by the \texttt{cleanTS()} function.
		\item
		\texttt{interval}: A string or numeric value, specifying the viewing resolution in the plot.
	\end{itemize}
	
	The problem with an animated plot is that the user does not have any control over the animation. There is not play or pause functionality so that the user can observe any desired frame. This can be achieved by adding interactivity to the plot. In the R programming language, \emph{shiny} \citep{shiny} provides a web application framework. It is an R package that creates interactive web apps using R. The \texttt{interact\_plot()} function creates and runs a shiny widget locally on the machine. It takes the \texttt{cleanTS} object and splits the cleaned data according to the \texttt{interval} argument, similar to the \texttt{animate\_interval()} function. It then creates a \emph{shiny widget} which shows the plot for the current state and gives a slider used to change the state. Unlike \texttt{animate\_interval()} it provides a global report containing information about complete data, and a state report giving information about the current state shown in the plot.\\
	
	%%%%%%%%%%%%%%%%%%%%
	
		%%%% FUNCTION 6 %%%%
	\item	\underline{\texttt{mergecsv()}} \label{mergecsv}
	
	\begin{Shaded}
\begin{Highlighting}[]
\FunctionTok{mergecsv}\NormalTok{(path, formats)}
\end{Highlighting}
	\end{Shaded}
	
	\begin{itemize}
		\tightlist
		\item
		\texttt{path}: The path to the folder containing the CSV files to merge.
		\item
		\texttt{formats}: The format of the timestamps used in the CSV files.
	\end{itemize}
	
	The \texttt{mergecsv()} function reads the CSV files found in the given path. It is assumed that in each CSV the first column contains the timestamps. All these files are read and the first column is parsed to a proper DateTime object using the formats given in the \texttt{formats} argument. Then these dataframes are merged using the timestamp column as a common column. The merged data frame returned by the function contains the first column as the timestamps. {The Appendix~\ref{apppendixA} demonstrates the working of this function with an example.}\\
	
	%%%%%%%%%%%%%%%%%%%%
	
\end{enumerate}

\subsection{The \emph{cleanTS} Web Application} \label{shinyapp}
{One of the requirements for using the package is having the R programming language installed on a local machine. Also, one needs to have at least some basic coding knowledge and experience to use the package. These drawbacks can be avoided using a web-based application, that runs on a web server and is accessed through a web browser. The user does not need to have R installed on their local systems. This enables users without any programming knowledge, to use the \emph{cleanTS} tool. The \emph{cleanTS} web app is created using \textit{shiny}, available at \url{https://mayur1009.shinyapps.io/cleanTS/}. The user needs to upload a CSV file containing the data and enter the format of the timestamps used in the data. The uploaded data and the statistical information of the data are calculated and displayed. The user then needs to select the imputation methods and press the start button. It is possible to add imputation methods by uploading an R source file containing the function. Once the data is cleaned, it is displayed along with its statistical information. The user can then download the cleaned data as a CSV file. The app also created an interactive plot for micro-scale visualization of the data. The plot can be converted to a GIF file and downloaded.}
%%%%%%%%%%%%%%%%%%%%%%%%%%%%%%%%%%%%%%%%%%

\section{Results and Conclusion}\label{conclusion}

Time series are used in a lot of different fields ranging from biology to social science and industries. The time series data is usually collected using sensors, which are prone to making errors and malfunctioning. These errors in data make the analytics unreliable. Bad analytics can greatly affect decision-making in businesses. Because the error rates are high, it has become very important to clean the time series data before using it. Also, it is found that data cleaning is a cumbersome task, especially in cases where the data is very large. A lot of time is consumed by the data cleaning process.

There are various tools developed for cleaning data, as discussed in the literature review of this paper. But many of these proposed tools are designed to operate on data for a specific field. Furthermore, many of them are not specifically designed for univariate time series data. A significant amount of research has been done related to missing value imputation. Missing data is a very common problem in real-world datasets, especially the ones recorded using sensors. Section~\ref{literature} discusses the various mechanisms of missing data. It also focuses on various tools and algorithms for missing data imputation.

This paper proposed a tool that automates the process of data cleaning for univariate time series data. The working and the implementation of the tool is also explained in detail. Firstly, the tool fixes any structural and datatype related errors. Then the timestamps of the data are observed for any missing timestamps or duplicate timestamps. Once the timestamps are fixed, missing values and anomalies or outliers in the data are handled. The tool also provides functions for visualizing the data in different resolutions. Section~\ref{examples} of this paper takes three different datasets to demonstrate the working of the proposed tool. Since the output plots generated are animated and interactive, they cannot be shown here.

\begin{table}[H]

\caption{\label{tab:perf}Measuring the running time of \emph{cleanTS} function.}
%\resizebox{\linewidth}{!}{
	\begin{tabular}{p{5em}rp{3em}rrp{3em}rl}
		\toprule
		\textbf{Data} & \textbf{Min} & \textbf{Lower Qrtl.} & \textbf{Mean} & \textbf{Median} & \textbf{Upper Qrtl.} & \textbf{Max} & \textbf{unit}\\
		\midrule
		\cellcolor{gray!6}{Power Consumption Dataset} & \cellcolor{gray!6}{19.99} & \cellcolor{gray!6}{20.87} & \cellcolor{gray!6}{21.66} & \cellcolor{gray!6}{21.57} & \cellcolor{gray!6}{22.29} & \cellcolor{gray!6}{24.59} & \cellcolor{gray!6}{sec}\\
		CO$_2$ Emission Dataset & 194.90 & 198.70 & 203.50 & 199.90 & 210.49 & 226.67 & msec\\
		\cellcolor{gray!6}{Temperature Dataset} & \cellcolor{gray!6}{13.49} & \cellcolor{gray!6}{14.18} & \cellcolor{gray!6}{14.38} & \cellcolor{gray!6}{14.42} & \cellcolor{gray!6}{14.56} & \cellcolor{gray!6}{15.10} & \cellcolor{gray!6}{sec}\\
		\bottomrule
	\end{tabular}
%}
\end{table}

The performance metric for the three datasets Power consumption, CO\textsubscript{2} emission, and temperature are listed in Table~\ref{tab:perf}. The power consumption dataset is very large and contains 121,273 observations. The CO\textsubscript{2} emission contains 1392 observations and the temperature dataset contains 45,253 observations. To evaluate the running time shown in Table \ref{tab:perf}, the \emph{microbenchmark} package \citep{microbenchmark} was used. From the examples shown in Section~\ref{examples}, it can be found that the package is user-friendly and easy to use. Also, the testing results show that the package is efficient and works well with a large amount of data. As of writing this paper, the package only works with univariate time series data, but it can be extended to work with multivariate datasets. Also, the package can be integrated to work with big data and databases by integration with Apache Spark. The package already works relatively well with a huge amount of data, but more efficiency can be achieved in future.

\subsection{Future Scope} \label{futurescope}
One of the limitations of the proposed package is that it only works with an univariate time series data. But it might be possible to add functionality that supports multivariate data, or even non-time series data. As of writing this article, the package uses the \texttt{data.table} library for working with the data. This is far more reliable and efficient at handling huge amounts of data than the base R dataframes. But it might be possible to integrate it with Apache Spark, to work with Big Data. It is also a worthwhile to investigate the possibility of adding parallel computing to make it faster. 

The tool will be modified for DNA and RNA sequencing applications, which have the potential to ease various processes involved in genetics and bioinformatics projects and experiments.

A major application of this tool will be in environmental datasets processing and analysis. The usability of the proposed tool and its GUI will be demonstrated for real-time energy market applications, which will automatically capture the time series, clean and organize it, analyze it and estimate the forecast with the least
possible error and generate a detailed report.

\noindent

% The description of your software  shall include:
% \begin{itemize}
% \item A short description of the high-level functionality and purpose of the software for a diverse, non-specialist audience
% \item An Impact overview that illustrates the purpose of the software and its achieved results:
% \begin{itemize}
% \item[-] Indicate in what way, and to what extent, the pursuit of existing research questions is improved. 
% \item[-] Indicate in what way new research questions can be pursued because of the software.
% \item[-] Indicate in what way the software has changed the daily practice of its users.
% \item[-] Indicate how widespread the use of the software is within and outside the intended user group.
% \item[-] Indicate in what way the software is used in commercial settings and/or how it led to the creation of spin-off companies (if so).
% \end{itemize}
% \item Mentions (if applicable) of any ongoing research projects using the software. 
% \item A list of all scholarly publications enabled by the software.
% \end{itemize}
\section*{Acknowledgements}
\label{acknowledgements}
The authors would like to thank Google Inc. for its support and funds to the project through Google Summer of Codes - 2021.
% \section*{References}

\appendix

\section{} \label{apppendixA}
{%
	This Appendix explains and gives an example for the \texttt{mergecsv()} function, explained in section~\ref{mergecsv}. We have four CSV files in the \textit{CSVFiles} folder. The first column of each of these files contains the timestamps. The formats argument contains a list of timestamp formats expected to be found while parsing the timestamp columns. 
}

\begin{Shaded}
\begin{Highlighting}[]
\CommentTok{\# Combine the csv files in the \textasciigrave{}CSVfiles\textasciigrave{} folder}
\NormalTok{merged }\OtherTok{\textless{}{-}} \FunctionTok{mergecsv}\NormalTok{(}\AttributeTok{path =} \StringTok{"CSVFiles/"}\NormalTok{, }
                    \AttributeTok{formats =} \FunctionTok{c}\NormalTok{(}\StringTok{"dmyHMs"}\NormalTok{, }\StringTok{"ymdHMS"}\NormalTok{))}

\CommentTok{\# Write the merged dataframe to \textasciigrave{}mergeCSV.csv\textasciigrave{} file}
\NormalTok{data.table}\SpecialCharTok{::}\FunctionTok{fwrite}\NormalTok{(merged, }\StringTok{"mergedCSV.csv"}\NormalTok{)}
\end{Highlighting}
\end{Shaded}

{%
	After the files are merged, the function returns a \texttt{data.table}, which can then be written to a CSV file using the \texttt{write.csv()} function or the \texttt{data.table::fwrite()} function.
}


%% References:

%\bibliographystyle{elsarticle-num} 
%\bibliography{references}



% % \bibitem{}

% \end{thebibliography}

\section*{Illustrative Examples} \label{examples}
% Optional : you may include one explanatory  video that will appear next to your article, in the right hand side panel. (Please upload any video as a single supplementary file with your article. Only one MP4 formatted, with 50MB maximum size, video is possible per article. Recommended video dimensions are 640 x 480 at a maximum of 30 frames / second. Prior to submission please test and validate your .mp4 file at  \url{http://elsevier-apps.sciverse.com/GadgetVideoPodcastPlayerWeb/verification} . This tool will display your video exactly in the same way as it will appear on ScienceDirect. )

\subsection*{Hourly Power Consumption}\label{ch4ex1}

 The data set used below is taken from Kaggle \citep{aepdata}. It contains over 10 years of hourly energy consumption data from 1st October 2004 to 3rd August 2018. The data is taken from PJM's website. PJM Interconnection LLC (PJM) is a regional transmission organization (RTO) in the United States. The hourly consumption data is recorded in megawatts(MW). There are 1,21,273 observations recorded in the dataset. The statistical information about the data is given in Table \ref{tab:ex1summary}.

\begin{Shaded}
\begin{Highlighting}[]
\CommentTok{\# Load the hourly data consumption data}
\NormalTok{data }\OtherTok{\textless{}{-}}\NormalTok{ data.table}\SpecialCharTok{::}\FunctionTok{fread}\NormalTok{(}\StringTok{"data/AEP\_hourly.csv"}\NormalTok{)}
\FunctionTok{summary}\NormalTok{(data)}
\end{Highlighting}
\end{Shaded}

\begin{verbatim}
##     Datetime                       AEP_MW
##  Min.   :2004-10-01 01:00:00   Min.   : 9581
##  1st Qu.:2008-03-17 15:00:00   1st Qu.:13630
##  Median :2011-09-02 04:00:00   Median :15310
##  Mean   :2011-09-02 03:17:01   Mean   :15500
##  3rd Qu.:2015-02-16 17:00:00   3rd Qu.:17200
##  Max.   :2018-08-03 00:00:00   Max.   :25695
\end{verbatim}

 \begin{table}[H]

 \caption{\label{tab:ex1summary}Statistical information of the Hourly consumption dataset.}
 \centering
 \begin{tabular}[t]{rrrrrr}
 \toprule
 \textbf{Min.} & \textbf{1st Qu.} & \textbf{Median} & \textbf{Mean} & \textbf{3rd Qu.} & \textbf{Max.}\\
 \midrule
 9581 & 13630 & 15310 & 15499.51 & 17200 & 25695\\
 \bottomrule
 \end{tabular}
 \end{table}

\begin{Shaded}
\begin{Highlighting}[]
\CommentTok{\# Load the cleanTS library}
\FunctionTok{library}\NormalTok{(cleanTS)}

\CommentTok{\# Use the \textasciigrave{}cleanTS()\textasciigrave{} function for cleaning the data.}
\NormalTok{cts }\OtherTok{\textless{}{-}} \FunctionTok{cleanTS}\NormalTok{(}\AttributeTok{data =}\NormalTok{ data, }\AttributeTok{date\_format =} \StringTok{"ymdHMs"}\NormalTok{, }
    \AttributeTok{replace\_outliers =}\NormalTok{ T)}
\end{Highlighting}
\end{Shaded}

\begin{Shaded}
\begin{Highlighting}[]
\CommentTok{\# The \textasciigrave{}cleanTS()\textasciigrave{} function returns a cleanTS object.}
\FunctionTok{summary}\NormalTok{(cts)}
\end{Highlighting}
\end{Shaded}

\begin{verbatim}
##                  Length Class      Mode
## clean_data        5     data.table list
## missing_ts       27     POSIXct    numeric
## duplicate_ts      4     POSIXct    numeric
## imp_methods       4     -none-     character
## mcar_err          4     data.frame list
## mar_err           0     data.frame list
## outliers          4     data.table list
## outlier_mcar_err  4     data.frame list
## outlier_mar_err   4     data.frame list
\end{verbatim}

\begin{Shaded}
\begin{Highlighting}[]
\CommentTok{\# Print the cleanTS object}
\FunctionTok{print}\NormalTok{(cts)}
\end{Highlighting}
\end{Shaded}

\begin{verbatim}
## $clean_data
## # A tibble: 121,296 x 5
##    time                value missing_type method_used is_outlier
##    <dttm>              <dbl> <chr>        <chr>       <lgl>
##  1 2004-10-01 01:00:00 12379 <NA>         <NA>        FALSE
##  2 2004-10-01 02:00:00 11935 <NA>         <NA>        FALSE
##  3 2004-10-01 03:00:00 11692 <NA>         <NA>        FALSE
##  4 2004-10-01 04:00:00 11597 <NA>         <NA>        FALSE
##  5 2004-10-01 05:00:00 11681 <NA>         <NA>        FALSE
##  6 2004-10-01 06:00:00 12280 <NA>         <NA>        FALSE
##  7 2004-10-01 07:00:00 13692 <NA>         <NA>        FALSE
##  8 2004-10-01 08:00:00 14618 <NA>         <NA>        FALSE
##  9 2004-10-01 09:00:00 14903 <NA>         <NA>        FALSE
## 10 2004-10-01 10:00:00 15118 <NA>         <NA>        FALSE
## # ... with 121,286 more rows
##
## $missing_ts
##  [1] "2004-10-31 02:00:00 UTC" "2005-04-03 03:00:00 UTC"
##  [3] "2005-10-30 02:00:00 UTC" "2006-04-02 03:00:00 UTC"
##  [5] "2006-10-29 02:00:00 UTC" "2007-03-11 03:00:00 UTC"
##  [7] "2007-11-04 02:00:00 UTC" "2008-03-09 03:00:00 UTC"
##  [9] "2008-11-02 02:00:00 UTC" "2009-03-08 03:00:00 UTC"
## [11] "2009-11-01 02:00:00 UTC" "2010-03-14 03:00:00 UTC"
## [13] "2010-11-07 02:00:00 UTC" "2010-12-10 00:00:00 UTC"
## [15] "2011-03-13 03:00:00 UTC" "2011-11-06 02:00:00 UTC"
## [17] "2012-03-11 03:00:00 UTC" "2012-11-04 02:00:00 UTC"
## [19] "2012-12-06 04:00:00 UTC" "2013-03-10 03:00:00 UTC"
## [21] "2013-11-03 02:00:00 UTC" "2014-03-09 03:00:00 UTC"
## [23] "2014-03-11 14:00:00 UTC" "2015-03-08 03:00:00 UTC"
## [25] "2016-03-13 03:00:00 UTC" "2017-03-12 03:00:00 UTC"
## [27] "2018-03-11 03:00:00 UTC"
##
## $duplicate_ts
## [1] "2014-11-02 02:00:00 UTC" "2015-11-01 02:00:00 UTC"
## [3] "2016-11-06 02:00:00 UTC" "2017-11-05 02:00:00 UTC"
##
## $imp_methods
## [1] "na_interpolation, na_locf, na_ma, na_kalman"
##
## $mcar_err
## # A tibble: 1 x 4
##   na_interpolation na_locf na_ma na_kalman
##              <dbl>   <dbl> <dbl>     <dbl>
## 1             2.84    9.17  6.45      1.84
##
## $mar_err
## # A tibble: 0 x 0
##
## $outliers
## # A tibble: 38 x 4
##    time                 value orig_value method_used
##    <dttm>               <dbl>      <dbl> <chr>
##  1 2006-05-30 16:00:00 22113.      22011 na_kalman
##  2 2006-05-30 17:00:00 22102.      22119 na_kalman
##  3 2007-07-09 15:00:00 23984.      23818 na_kalman
##  4 2007-07-09 16:00:00 24229.      23940 na_kalman
##  5 2007-07-09 17:00:00 24236.      24038 na_kalman
##  6 2007-08-23 16:00:00 24974.      24828 na_kalman
##  7 2007-08-23 17:00:00 24984.      24862 na_kalman
##  8 2008-06-09 15:00:00 24077.      23938 na_kalman
##  9 2008-06-09 16:00:00 24145.      23828 na_kalman
## 10 2008-06-09 17:00:00 24005.      23900 na_kalman
## # ... with 28 more rows
##
## $outlier_mcar_err
## # A tibble: 1 x 4
##   na_interpolation na_locf na_ma na_kalman
##              <dbl>   <dbl> <dbl>     <dbl>
## 1            0.726    2.67  1.97     0.340
##
## $outlier_mar_err
## # A tibble: 1 x 4
##   na_interpolation na_locf na_ma na_kalman
##              <dbl>   <dbl> <dbl>     <dbl>
## 1             30.6    43.2  32.1      27.3
\end{verbatim}

\begin{Shaded}
\begin{Highlighting}[]
\CommentTok{\# Use the \textasciigrave{}gen.report()\textasciigrave{} function to get a detailed report.}
\FunctionTok{gen.report}\NormalTok{(cts)}
\end{Highlighting}
\end{Shaded}

\begin{verbatim}
##
## # Summary of cleaned data:
##    Min. 1st Qu.  Median    Mean 3rd Qu.    Max.
##    9581   13629   15309   15499   17200   25164
##
## # Missing timestamps:  27
##  [1] "2004-10-31 02:00:00 UTC" "2005-04-03 03:00:00 UTC"
##  [3] "2005-10-30 02:00:00 UTC" "2006-04-02 03:00:00 UTC"
##  [5] "2006-10-29 02:00:00 UTC" "2007-03-11 03:00:00 UTC"
##  [7] "2007-11-04 02:00:00 UTC" "2008-03-09 03:00:00 UTC"
##  [9] "2008-11-02 02:00:00 UTC" "2009-03-08 03:00:00 UTC"
## [11] "2009-11-01 02:00:00 UTC" "2010-03-14 03:00:00 UTC"
## [13] "2010-11-07 02:00:00 UTC" "2010-12-10 00:00:00 UTC"
## [15] "2011-03-13 03:00:00 UTC" "2011-11-06 02:00:00 UTC"
## [17] "2012-03-11 03:00:00 UTC" "2012-11-04 02:00:00 UTC"
## [19] "2012-12-06 04:00:00 UTC" "2013-03-10 03:00:00 UTC"
## [21] "2013-11-03 02:00:00 UTC" "2014-03-09 03:00:00 UTC"
## [23] "2014-03-11 14:00:00 UTC" "2015-03-08 03:00:00 UTC"
## [25] "2016-03-13 03:00:00 UTC" "2017-03-12 03:00:00 UTC"
## [27] "2018-03-11 03:00:00 UTC"
##
## # Duplicate timestamps:  0
##
## No duplicate timestamps found.
##
## # Missing Values:  31 (0.0255573143384778%)
##
## ## MCAR:  31 (0.0255573143384778%)
##  MCAR Errors:
##   na_interpolation  na_locf    na_ma na_kalman
## 1         2.844675 9.173093 6.446093  1.836841
##
##                    time    value method_used
##  1: 2004-10-31 02:00:00 10759.00   na_kalman
##  2: 2005-04-03 03:00:00 13334.83   na_kalman
##  3: 2005-10-30 02:00:00 13158.50   na_kalman
##  4: 2006-04-02 03:00:00 11234.17   na_kalman
##  5: 2006-10-29 02:00:00 13128.33   na_kalman
##  6: 2007-03-11 03:00:00 13017.33   na_kalman
##  7: 2007-11-04 02:00:00 13347.50   na_kalman
##  8: 2008-03-09 03:00:00 17156.83   na_kalman
##  9: 2008-11-02 02:00:00 12292.50   na_kalman
## 10: 2009-03-08 03:00:00 10991.67   na_kalman
## 11: 2009-11-01 02:00:00 11551.00   na_kalman
## 12: 2010-03-14 03:00:00 12546.00   na_kalman
## 13: 2010-11-07 02:00:00 14467.67   na_kalman
## 14: 2010-12-10 00:00:00 18107.33   na_kalman
## 15: 2011-03-13 03:00:00 12787.83   na_kalman
## 16: 2011-11-06 02:00:00 13279.33   na_kalman
## 17: 2012-03-11 03:00:00 13397.50   na_kalman
## 18: 2012-11-04 02:00:00 12432.17   na_kalman
## 19: 2012-12-06 04:00:00 14801.50   na_kalman
## 20: 2013-03-10 03:00:00 12429.83   na_kalman
## 21: 2013-11-03 02:00:00 11713.33   na_kalman
## 22: 2014-03-09 03:00:00 13021.00   na_kalman
## 23: 2014-03-11 14:00:00 14635.33   na_kalman
## 24: 2014-11-02 02:00:00 12952.17   na_kalman
## 25: 2015-03-08 03:00:00 14044.50   na_kalman
## 26: 2015-11-01 02:00:00 10695.50   na_kalman
## 27: 2016-03-13 03:00:00 10218.17   na_kalman
## 28: 2016-11-06 02:00:00 11049.83   na_kalman
## 29: 2017-03-12 03:00:00 14301.83   na_kalman
## 30: 2017-11-05 02:00:00 10535.67   na_kalman
## 31: 2018-03-11 03:00:00 13722.17   na_kalman
##                    time    value method_used
##
##
## ## MAR:  0 (0%)
## No MAR found.
##
## # Outliers:  38
##                    time    value orig_value method_used
##  1: 2006-05-30 16:00:00 22112.80      22011   na_kalman
##  2: 2006-05-30 17:00:00 22101.70      22119   na_kalman
##  3: 2007-07-09 15:00:00 23984.40      23818   na_kalman
##  4: 2007-07-09 16:00:00 24228.80      23940   na_kalman
##  5: 2007-07-09 17:00:00 24235.80      24038   na_kalman
##  6: 2007-08-23 16:00:00 24974.40      24828   na_kalman
##  7: 2007-08-23 17:00:00 24984.10      24862   na_kalman
##  8: 2008-06-09 15:00:00 24077.40      23938   na_kalman
##  9: 2008-06-09 16:00:00 24145.30      23828   na_kalman
## 10: 2008-06-09 17:00:00 24004.80      23900   na_kalman
## 11: 2008-10-20 14:00:00 16174.83      25695   na_kalman
## 12: 2009-03-03 07:00:00 21756.17      22068   na_kalman
## 13: 2010-08-30 16:00:00 22847.60      22777   na_kalman
## 14: 2010-08-30 17:00:00 22898.40      22958   na_kalman
## 15: 2010-08-31 16:00:00 22789.80      22839   na_kalman
## 16: 2010-08-31 17:00:00 22989.20      23023   na_kalman
## 17: 2011-09-02 15:00:00 22741.00      22666   na_kalman
## 18: 2011-09-02 16:00:00 22967.00      22826   na_kalman
## 19: 2011-09-02 17:00:00 22899.00      22893   na_kalman
## 20: 2012-06-30 08:00:00 10021.30      10015   na_kalman
## 21: 2012-06-30 09:00:00 10588.20      10582   na_kalman
## 22: 2013-09-10 14:00:00 21876.21      22016   na_kalman
## 23: 2013-09-10 15:00:00 22391.00      22631   na_kalman
## 24: 2013-09-10 16:00:00 22630.71      22781   na_kalman
## 25: 2013-09-10 17:00:00 22611.71      22722   na_kalman
## 26: 2013-09-10 18:00:00 22350.36      22433   na_kalman
## 27: 2014-01-07 01:00:00 21879.29      21807   na_kalman
## 28: 2014-01-07 02:00:00 21893.30      21684   na_kalman
## 29: 2014-01-07 03:00:00 22048.44      21689   na_kalman
## 30: 2014-01-07 04:00:00 22300.15      21785   na_kalman
## 31: 2014-01-07 05:00:00 22603.85      21892   na_kalman
## 32: 2014-01-07 06:00:00 22914.96      22278   na_kalman
## 33: 2014-01-07 07:00:00 23188.90      23076   na_kalman
## 34: 2014-01-07 08:00:00 23381.11      23590   na_kalman
## 35: 2015-01-08 05:00:00 21587.60      21435   na_kalman
## 36: 2015-01-08 06:00:00 22393.90      22182   na_kalman
## 37: 2015-01-08 07:00:00 23173.00      23056   na_kalman
## 38: 2015-02-20 06:00:00 23397.83      23412   na_kalman
##                    time    value orig_value method_used
## ## Imputation errors while replacing outliers:
## ### MCAR errors:
##   na_interpolation  na_locf    na_ma na_kalman
## 1        0.7262274 2.670749 1.974189 0.3398252
## ### MAR errors:
##   na_interpolation  na_locf    na_ma na_kalman
## 1         30.59071 43.16642 32.12617  27.31822
\end{verbatim}

\begin{Shaded}
\begin{Highlighting}[]
\CommentTok{\# Use the \textasciigrave{}animate\_interval()\textasciigrave{} function to create a animation}
\CommentTok{\# object.}
\NormalTok{anim }\OtherTok{\textless{}{-}} \FunctionTok{animate\_interval}\NormalTok{(cts, }\AttributeTok{interval =} \StringTok{"1 month"}\NormalTok{)}
\end{Highlighting}
\end{Shaded}

\begin{Shaded}
\begin{Highlighting}[]
\CommentTok{\# The animation is generated using \textasciigrave{}gen.animation()\textasciigrave{} function.}
\FunctionTok{gen.animation}\NormalTok{(anim)}
\CommentTok{\# This animation can be saved using \textasciigrave{}anim\_save()\textasciigrave{} function.}
\end{Highlighting}
\end{Shaded}

     \begin{figure}[htbp]
     {
       \centering
       \includegraphics[height = \textwidth, angle = 270]{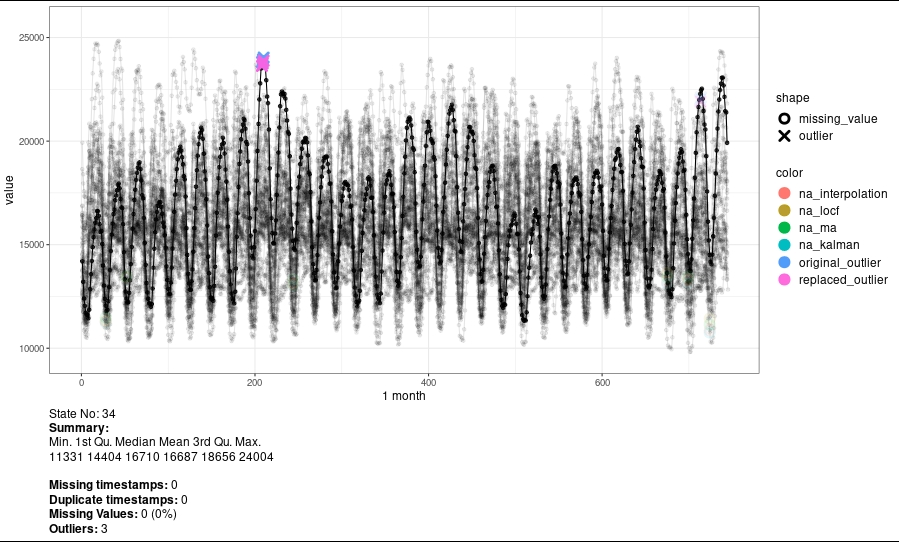}
     }
     \caption{A state from the animated plot generated for Example 1}\label{fig:ex1gif}
     \end{figure}

\begin{Shaded}
\begin{Highlighting}[]
\CommentTok{\# Start a interactive plot using the \textasciigrave{}interact\_plot()\textasciigrave{}}
\CommentTok{\# function.}
\FunctionTok{interact\_plot}\NormalTok{(cts, }\AttributeTok{interval =} \StringTok{"1 month"}\NormalTok{)}
\end{Highlighting}
\end{Shaded}

     \begin{figure}[htbp]
     {
       \centering
       \includegraphics[height = \textwidth, angle = 270]{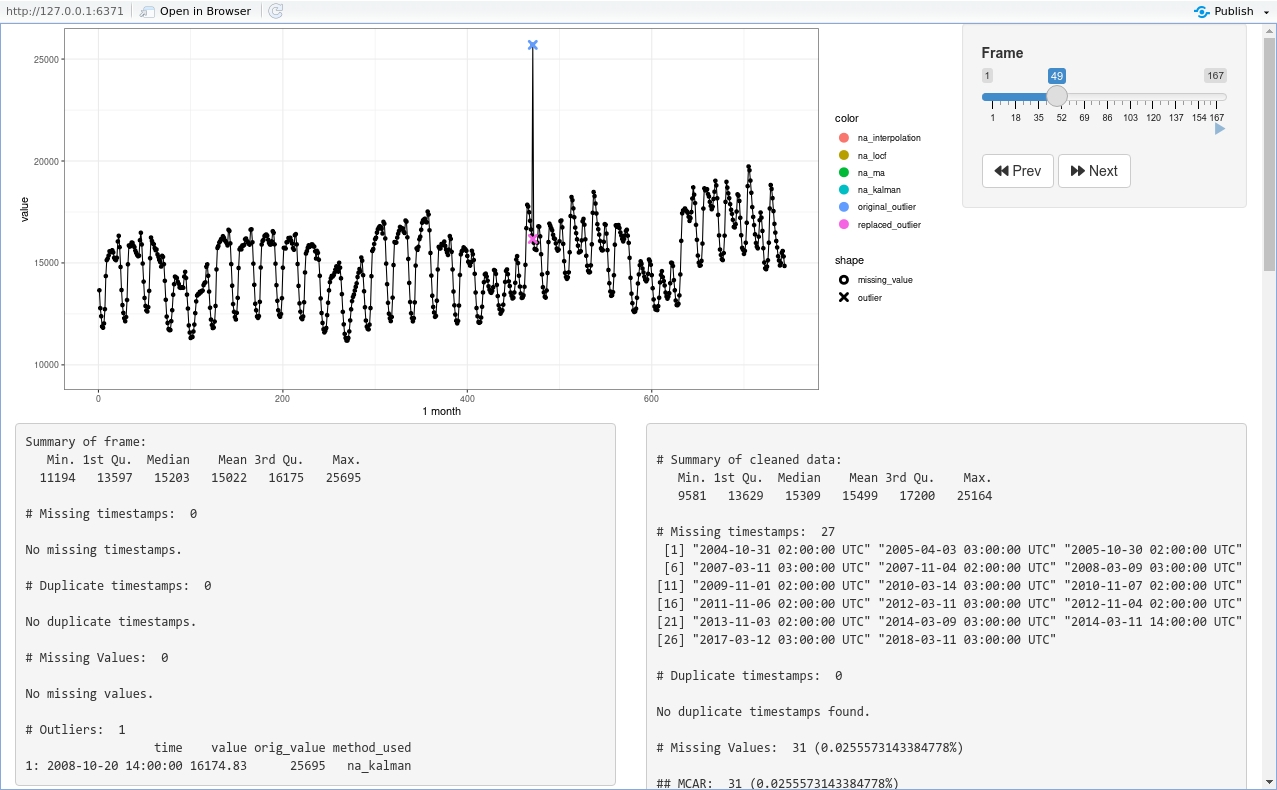}
     }
     \caption{A state from the interactive plot generated for Example 1}\label{fig:ex1intp}
     \end{figure}

 Figure \ref{fig:ex1gif} and Figure \ref{fig:ex1intp} shows a single state from the animated and interactive plots respectively.

 \subsection*{Carbon-dioxide Emission}\label{ch4ex2}

\begin{figure}[htbp]

 {\centering \includegraphics[width=1\linewidth,]{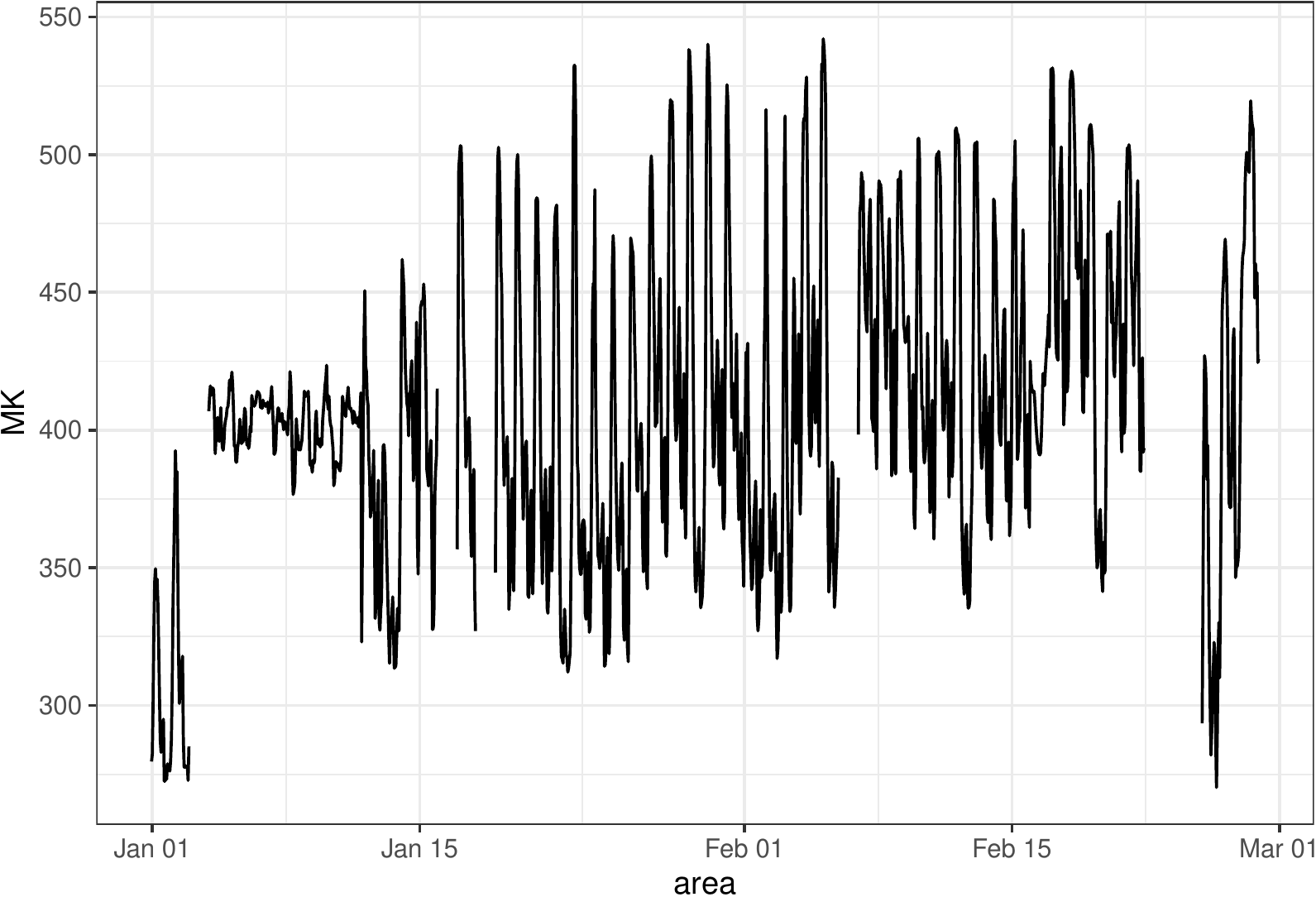}

 }

 \caption{Plot for the CO2 emission dataset}\label{fig:ex2completeplot}
 \end{figure}

 For this example, the CO2 emission dataset is used. This dataset was used in \citep{33,bokde2021optimal}, for short-term CO\textsubscript{2} emission forecasting. Data is used from the ENSTO-E transparency platform for 2018 and 2019. The dataset contains the electricity generation per technology, electricity demand per price area, as well as power, flows between interconnected areas. The flow tracing is used to map the power flows between importing and exporting countries. The country-specific average CO\textsubscript{2} emission intensity per generation technology is applied to the flow tracing results to calculate the hourly CO\textsubscript{2} intensity of electricity consumption for each price area. The statistical information for this dataset is shown in Table \ref{tab:ex2summary}. Figure \ref{fig:ex2completeplot} shows the plot for the dataset.

 \begin{table}[H]

 \caption{\label{tab:ex2summary}Statistical information of the CO2 emission dataset.}
 \centering
 \begin{tabular}[t]{rrrrrr}
 \toprule
 \textbf{Min.} & \textbf{1st Qu.} & \textbf{Median} & \textbf{Mean} & \textbf{3rd Qu.} & \textbf{Max.}\\
 \midrule
 270.23 & 370.46 & 402.545 & 405.9396 & 437.5025 & 542.06\\
 \bottomrule
 \end{tabular}
 \end{table}

\begin{Shaded}
\begin{Highlighting}[]
\CommentTok{\# Load the hourly data consumption data}
\NormalTok{data }\OtherTok{\textless{}{-}}\NormalTok{ data.table}\SpecialCharTok{::}\FunctionTok{fread}\NormalTok{(}\StringTok{"data/co2dat.csv"}\NormalTok{)}
\FunctionTok{summary}\NormalTok{(data)}
\end{Highlighting}
\end{Shaded}

\begin{verbatim}
##       area                           MK
##  Min.   :2016-12-31 23:00:00   Min.   :270.2
##  1st Qu.:2017-01-15 10:45:00   1st Qu.:370.5
##  Median :2017-01-29 22:30:00   Median :402.5
##  Mean   :2017-01-29 22:30:00   Mean   :405.9
##  3rd Qu.:2017-02-13 10:15:00   3rd Qu.:437.5
##  Max.   :2017-02-27 22:00:00   Max.   :542.1
##                                NA's   :168
\end{verbatim}

\begin{Shaded}
\begin{Highlighting}[]
\CommentTok{\# Load the cleanTS library}
\FunctionTok{library}\NormalTok{(cleanTS)}

\CommentTok{\# Use the \textasciigrave{}cleanTS()\textasciigrave{} function for}
\CommentTok{\# cleaning the data.}
\NormalTok{cts }\OtherTok{\textless{}{-}} \FunctionTok{cleanTS}\NormalTok{(}\AttributeTok{data =}\NormalTok{ data, }\AttributeTok{date\_format =} \StringTok{"ymdHMs"}\NormalTok{, }
    \AttributeTok{replace\_outliers =}\NormalTok{ T)}
\end{Highlighting}
\end{Shaded}

\begin{Shaded}
\begin{Highlighting}[]
\CommentTok{\# The \textasciigrave{}cleanTS()\textasciigrave{} function returns a cleanTS object.}
\FunctionTok{summary}\NormalTok{(cts)}
\end{Highlighting}
\end{Shaded}

\begin{verbatim}
##                  Length Class      Mode
## clean_data       5      data.table list
## missing_ts       0      POSIXct    numeric
## duplicate_ts     0      POSIXct    numeric
## imp_methods      4      -none-     character
## mcar_err         0      data.frame list
## mar_err          4      data.frame list
## outliers         4      data.table list
## outlier_mcar_err 0      data.frame list
## outlier_mar_err  0      data.frame list
\end{verbatim}

\begin{Shaded}
\begin{Highlighting}[]
\CommentTok{\# Print the cleanTS object}
\FunctionTok{print}\NormalTok{(cts)}
\end{Highlighting}
\end{Shaded}

\begin{verbatim}
## $clean_data
## # A tibble: 1,392 x 5
##    time                value missing_type method_used is_outlier
##    <dttm>              <dbl> <chr>        <chr>       <lgl>
##  1 2016-12-31 23:00:00  280. <NA>         <NA>        FALSE
##  2 2017-01-01 00:00:00  282. <NA>         <NA>        FALSE
##  3 2017-01-01 01:00:00  301. <NA>         <NA>        FALSE
##  4 2017-01-01 02:00:00  328. <NA>         <NA>        FALSE
##  5 2017-01-01 03:00:00  345. <NA>         <NA>        FALSE
##  6 2017-01-01 04:00:00  350. <NA>         <NA>        FALSE
##  7 2017-01-01 05:00:00  340. <NA>         <NA>        FALSE
##  8 2017-01-01 06:00:00  346. <NA>         <NA>        FALSE
##  9 2017-01-01 07:00:00  338. <NA>         <NA>        FALSE
## 10 2017-01-01 08:00:00  322. <NA>         <NA>        FALSE
## # ... with 1,382 more rows
##
## $missing_ts
## POSIXct of length 0
##
## $duplicate_ts
## POSIXct of length 0
##
## $imp_methods
## [1] "na_interpolation, na_locf, na_ma, na_kalman"
##
## $mcar_err
## # A tibble: 0 x 0
##
## $mar_err
## # A tibble: 1 x 4
##   na_interpolation na_locf na_ma na_kalman
##              <dbl>   <dbl> <dbl>     <dbl>
## 1             20.0    21.8  22.0      86.0
##
## $outliers
## # A tibble: 0 x 4
## # ... with 4 variables: time <dttm>, value <dbl>, method_used <lgl>,
## #   orig_value <dbl>
##
## $outlier_mcar_err
## # A tibble: 0 x 0
##
## $outlier_mar_err
## # A tibble: 0 x 0
\end{verbatim}

\begin{Shaded}
\begin{Highlighting}[]
\CommentTok{\# Use the \textasciigrave{}gen.report()\textasciigrave{} function to get a detailed report.}
\FunctionTok{gen.report}\NormalTok{(cts)}
\end{Highlighting}
\end{Shaded}

\begin{verbatim}
##
## # Summary of cleaned data:
##    Min. 1st Qu.  Median    Mean 3rd Qu.    Max.
##   270.2   362.0   396.4   399.9   430.0   542.1
##
## # Missing timestamps:  0
##
## No missing timestamps found.
##
## # Duplicate timestamps:  0
##
## No duplicate timestamps found.
##
## # Missing Values:  168 (12.0689655172414%)
##
## ## MCAR:  0 (0%)
## No MCAR found.
##
##
## ## MAR:  168 (12.0689655172414%)
##  MAR Errors:
##   na_interpolation  na_locf    na_ma na_kalman
## 1         19.96321 21.83961 21.95224  85.97244
##                     time    value      method_used
##   1: 2017-01-02 23:00:00 289.9804 na_interpolation
##   2: 2017-01-03 00:00:00 294.8508 na_interpolation
##   3: 2017-01-03 01:00:00 299.7212 na_interpolation
##   4: 2017-01-03 02:00:00 304.5916 na_interpolation
##   5: 2017-01-03 03:00:00 309.4620 na_interpolation
##  ---
## 164: 2017-02-24 18:00:00 300.3266 na_interpolation
## 165: 2017-02-24 19:00:00 298.9573 na_interpolation
## 166: 2017-02-24 20:00:00 297.5879 na_interpolation
## 167: 2017-02-24 21:00:00 296.2186 na_interpolation
## 168: 2017-02-24 22:00:00 294.8493 na_interpolation
##
## # Outliers:  0
##
## No outliers found.
\end{verbatim}

 \subsection*{Temperature Dataset}\label{ch4ex3}

\begin{figure}[htbp]

 {\centering \includegraphics[width=1\linewidth,]{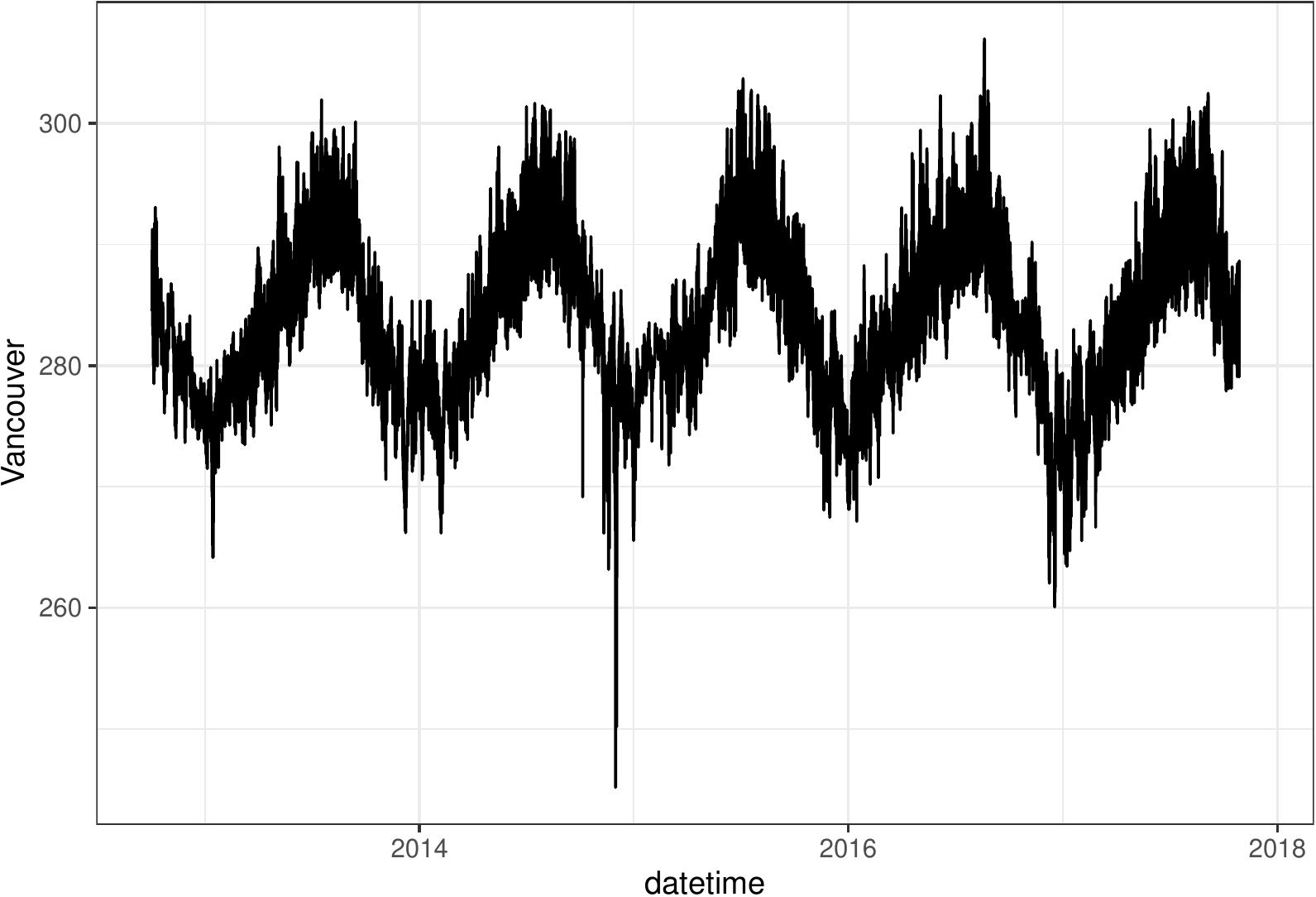}

 }

 \caption{Plot for the Temperature dataset in Example 3.}\label{fig:ex3completeplot}
 \end{figure}

 \citep{tempdata} provides a dataset of high temporal resolution(hourly measurement) data of weather attributes, like, humidity, air pressure, wind speed, wind direction, temperature, etc. It contains data for approximately 5 years and 36 different cities. For this example, the temperature data for Vancouver city is used. The recorded observations in the data are in Kelvin. Figure \ref{fig:ex3completeplot} and Table \ref{tab:ex3summary} show the plot and the summary of the data being used.

 \begin{table}[H]

 \caption{\label{tab:ex3summary}Statistical information of the Temperature dataset in Example 3.}
 \centering
 \begin{tabular}[t]{rrrrrr}
 \toprule
 \textbf{Min.} & \textbf{1st Qu.} & \textbf{Median} & \textbf{Mean} & \textbf{3rd Qu.} & \textbf{Max.}\\
 \midrule
 245.15 & 279.16 & 283.45 & 283.8627 & 288.6008 & 307\\
 \bottomrule
 \end{tabular}
 \end{table}

\begin{Shaded}
\begin{Highlighting}[]
\CommentTok{\# Load the hourly data consumption data}
\NormalTok{data }\OtherTok{\textless{}{-}}\NormalTok{ data.table}\SpecialCharTok{::}\FunctionTok{fread}\NormalTok{(}\StringTok{"data/temperature.csv"}\NormalTok{)}
\FunctionTok{summary}\NormalTok{(data[, }\FunctionTok{c}\NormalTok{(}\StringTok{"datetime"}\NormalTok{, }\StringTok{"Vancouver"}\NormalTok{)])}
\end{Highlighting}
\end{Shaded}

\begin{verbatim}
##     datetime                     Vancouver
##  Min.   :2012-10-01 12:00:00   Min.   :245.2
##  1st Qu.:2014-01-15 21:00:00   1st Qu.:279.2
##  Median :2015-05-02 06:00:00   Median :283.4
##  Mean   :2015-05-02 06:00:00   Mean   :283.9
##  3rd Qu.:2016-08-15 15:00:00   3rd Qu.:288.6
##  Max.   :2017-11-30 00:00:00   Max.   :307.0
##                                NA's   :795
\end{verbatim}

\begin{Shaded}
\begin{Highlighting}[]
\CommentTok{\# Load the cleanTS library}
\FunctionTok{library}\NormalTok{(cleanTS)}

\CommentTok{\# Use the \textasciigrave{}cleanTS()\textasciigrave{} function for cleaning the data.}
\NormalTok{cts }\OtherTok{\textless{}{-}} \FunctionTok{cleanTS}\NormalTok{(}\AttributeTok{data =}\NormalTok{ data, }\AttributeTok{date\_format =} \StringTok{"ymdHMs"}\NormalTok{, }
    \AttributeTok{time =} \StringTok{"datetime"}\NormalTok{, }\AttributeTok{value =} \StringTok{"Vancouver"}\NormalTok{)}

\CommentTok{\# The dataset contains multiples columns, so the time and}
\CommentTok{\# value argument is used to specify the \textasciigrave{}timestamp\textasciigrave{} and to}
\CommentTok{\# specify the \textasciigrave{}timestamp\textasciigrave{} and \textasciigrave{}observation\textasciigrave{} columns manually.}
\end{Highlighting}
\end{Shaded}

\begin{Shaded}
\begin{Highlighting}[]
\CommentTok{\# The \textasciigrave{}cleanTS()\textasciigrave{} function returns a cleanTS object.}
\FunctionTok{summary}\NormalTok{(cts)}
\end{Highlighting}
\end{Shaded}

\begin{verbatim}
##                  Length Class      Mode
## clean_data       5      data.table list
## missing_ts       0      POSIXct    numeric
## duplicate_ts     0      POSIXct    numeric
## imp_methods      4      -none-     character
## mcar_err         4      data.frame list
## mar_err          4      data.frame list
## outliers         4      data.table list
## outlier_mcar_err 4      data.frame list
## outlier_mar_err  4      data.frame list
\end{verbatim}

\begin{Shaded}
\begin{Highlighting}[]
\CommentTok{\# Print the cleanTS object}
\FunctionTok{print}\NormalTok{(cts)}
\end{Highlighting}
\end{Shaded}

\begin{verbatim}
## $clean_data
## # A tibble: 45,253 x 5
##    time                value missing_type method_used      is_outlier
##    <dttm>              <dbl> <chr>        <chr>            <lgl>
##  1 2012-10-01 12:00:00  285. mcar         na_interpolation FALSE
##  2 2012-10-01 13:00:00  285. <NA>         <NA>             FALSE
##  3 2012-10-01 14:00:00  285. <NA>         <NA>             FALSE
##  4 2012-10-01 15:00:00  285. <NA>         <NA>             FALSE
##  5 2012-10-01 16:00:00  285. <NA>         <NA>             FALSE
##  6 2012-10-01 17:00:00  285. <NA>         <NA>             FALSE
##  7 2012-10-01 18:00:00  285. <NA>         <NA>             FALSE
##  8 2012-10-01 19:00:00  285. <NA>         <NA>             FALSE
##  9 2012-10-01 20:00:00  285. <NA>         <NA>             FALSE
## 10 2012-10-01 21:00:00  285. <NA>         <NA>             FALSE
## # ... with 45,243 more rows
##
## $missing_ts
## POSIXct of length 0
##
## $duplicate_ts
## POSIXct of length 0
##
## $imp_methods
## [1] "na_interpolation, na_locf, na_ma, na_kalman"
##
## $mcar_err
## # A tibble: 1 x 4
##   na_interpolation na_locf   na_ma na_kalman
##              <dbl>   <dbl>   <dbl>     <dbl>
## 1          0.00160 0.00258 0.00213   0.00164
##
## $mar_err
## # A tibble: 1 x 4
##   na_interpolation na_locf na_ma na_kalman
##              <dbl>   <dbl> <dbl>     <dbl>
## 1            0.493   0.584 0.571      3.09
##
## $outliers
## # A tibble: 33 x 4
##    time                value orig_value method_used
##    <dttm>              <dbl>      <dbl> <chr>
##  1 2014-10-05 18:00:00  290.       269. na_ma
##  2 2014-11-10 18:00:00  279.       266. na_kalman
##  3 2014-11-10 19:00:00  281.       267. na_kalman
##  4 2014-11-10 20:00:00  282.       270. na_kalman
##  5 2014-11-18 19:00:00  271.       263. na_kalman
##  6 2014-11-18 20:00:00  271.       267. na_kalman
##  7 2014-11-20 08:00:00  275.       265. na_ma
##  8 2014-11-29 22:00:00  271.       271. na_kalman
##  9 2014-11-29 23:00:00  270.       271. na_kalman
## 10 2014-11-30 00:00:00  270.       270. na_kalman
## # ... with 23 more rows
##
## $outlier_mcar_err
## # A tibble: 1 x 4
##   na_interpolation na_locf   na_ma na_kalman
##              <dbl>   <dbl>   <dbl>     <dbl>
## 1          0.00697  0.0122 0.00648   0.00704
##
## $outlier_mar_err
## # A tibble: 1 x 4
##   na_interpolation na_locf  na_ma na_kalman
##              <dbl>   <dbl>  <dbl>     <dbl>
## 1           0.0659  0.0974 0.0708    0.0471
\end{verbatim}

\begin{Shaded}
\begin{Highlighting}[]
\CommentTok{\# Use the \textasciigrave{}gen.report()\textasciigrave{} function to get a detailed report.}
\FunctionTok{gen.report}\NormalTok{(cts)}
\end{Highlighting}
\end{Shaded}

\begin{verbatim}
##
## # Summary of cleaned data:
##    Min. 1st Qu.  Median    Mean 3rd Qu.    Max.
##   260.1   279.2   283.6   283.9   288.5   305.4
##
## # Missing timestamps:  0
##
## No missing timestamps found.
##
## # Duplicate timestamps:  0
##
## No duplicate timestamps found.
##
## # Missing Values:  795 (1.75678960510905%)
##
## ## MCAR:  1 (0.00220979824542019%)
##  MCAR Errors:
##   na_interpolation     na_locf       na_ma   na_kalman
## 1      0.001603655 0.002577279 0.002134501 0.001635805
##
##                   time  value      method_used
## 1: 2012-10-01 12:00:00 284.63 na_interpolation
##
##
## ## MAR:  794 (1.75457980686363%)
##  MAR Errors:
##   na_interpolation  na_locf     na_ma na_kalman
## 1        0.4931633 0.583915 0.5714428  3.094038
##                     time    value      method_used
##   1: 2013-03-11 07:00:00 278.6433 na_interpolation
##   2: 2013-03-11 08:00:00 278.5267 na_interpolation
##   3: 2017-10-28 01:00:00 288.0100 na_interpolation
##   4: 2017-10-28 02:00:00 288.0100 na_interpolation
##   5: 2017-10-28 03:00:00 288.0100 na_interpolation
##  ---
## 790: 2017-11-29 20:00:00 288.0100 na_interpolation
## 791: 2017-11-29 21:00:00 288.0100 na_interpolation
## 792: 2017-11-29 22:00:00 288.0100 na_interpolation
## 793: 2017-11-29 23:00:00 288.0100 na_interpolation
## 794: 2017-11-30 00:00:00 288.0100 na_interpolation
##
## # Outliers:  33
##                    time    value orig_value method_used
##  1: 2014-10-05 18:00:00 289.7553   269.1500       na_ma
##  2: 2014-11-10 18:00:00 279.3756   266.1500   na_kalman
##  3: 2014-11-10 19:00:00 280.7293   267.1500   na_kalman
##  4: 2014-11-10 20:00:00 281.9608   269.7883   na_kalman
##  5: 2014-11-18 19:00:00 270.9114   263.1500   na_kalman
##  6: 2014-11-18 20:00:00 271.0321   266.6466   na_kalman
##  7: 2014-11-20 08:00:00 274.9799   264.8900       na_ma
##  8: 2014-11-29 22:00:00 270.5659   270.5791   na_kalman
##  9: 2014-11-29 23:00:00 270.4940   270.5400   na_kalman
## 10: 2014-11-30 00:00:00 270.3393   270.3260   na_kalman
## 11: 2014-11-30 01:00:00 270.1110   269.8300   na_kalman
## 12: 2014-11-30 02:00:00 269.8186   269.5400   na_kalman
## 13: 2014-11-30 08:00:00 268.1006   266.8008       na_ma
## 14: 2014-11-30 16:00:00 267.4040   266.3400   na_kalman
## 15: 2014-11-30 17:00:00 268.0969   266.3906   na_kalman
## 16: 2014-11-30 18:00:00 268.8950   258.7303   na_kalman
## 17: 2014-11-30 19:00:00 269.7435   245.1500   na_kalman
## 18: 2014-11-30 20:00:00 270.5876   248.9412   na_kalman
## 19: 2014-11-30 21:00:00 271.3725   254.9672   na_kalman
## 20: 2014-12-02 16:00:00 270.0522   266.9844       na_ma
## 21: 2014-12-02 18:00:00 271.7652   267.3107   na_kalman
## 22: 2014-12-02 19:00:00 273.3105   250.1500   na_kalman
## 23: 2016-06-06 00:00:00 301.6683   302.3100   na_kalman
## 24: 2016-06-06 01:00:00 301.3838   301.9200   na_kalman
## 25: 2016-06-06 02:00:00 300.8096   301.3100   na_kalman
## 26: 2016-08-19 21:00:00 305.1987   305.9000   na_kalman
## 27: 2016-08-19 22:00:00 305.3927   306.6900   na_kalman
## 28: 2016-08-19 23:00:00 305.2390   307.0000   na_kalman
## 29: 2016-08-20 00:00:00 304.7360   306.6900   na_kalman
## 30: 2016-08-20 01:00:00 303.8823   306.0600   na_kalman
## 31: 2016-08-20 02:00:00 302.6764   305.3000   na_kalman
## 32: 2016-12-18 03:00:00 265.3639   262.8000   na_kalman
## 33: 2016-12-18 04:00:00 263.9526   262.5300   na_kalman
##                    time    value orig_value method_used
## ## Imputation errors while replacing outliers:
## ### MCAR errors:
##   na_interpolation    na_locf       na_ma   na_kalman
## 1      0.006966497 0.01220904 0.006478862 0.007041101
## ### MAR errors:
##   na_interpolation    na_locf      na_ma  na_kalman
## 1       0.06589995 0.09743086 0.07080976 0.04711339
\end{verbatim}

 The output generated by the \texttt{animate\_interval()} is a GIF, and therefore it is not possible to include it here. Similarly, the output of \texttt{interactive\_plot()} is a interactive object cannot be shown. These outputs for all the examples shown here are available at this link: \href{https://drive.google.com/drive/folders/1NYvHcib2JGDgPSyN_3uAYOBO6Dv8N6di?usp=sharing}{\textbf{\textit{\underline{Link to Outputs}}}} (\emph{\url{https://drive.google.com/drive/folders/1NYvHcib2JGDgPSyN_3uAYOBO6Dv8N6di?usp=sharing}}).

%\section*{Required Metadata}
%\label{reqmeta}

\section*{Current code version}
\label{curcodever}

%Ancillary data table required for subversion of the codebase. Kindly replace examples in right column with the correct information about your current code, and leave the left column as it is.

\begin{table}[H]
\begin{tabular}{|l|p{5cm}|p{9cm}|}
\hline
\textbf{Nr.} & \textbf{Code metadata description} & \textbf{Please fill in this column} \\
\hline
C1 & Current code version & v0.1.0 \\
\hline
C2 & Permanent link to code/repository used for this code version & $https://github.com/Mayur1009/cleanTS$, $https://cran.r-project.org/package=cleanTS$\\
\hline
C3  & Permanent link to Reproducible Capsule & \\
\hline
C4 & Legal Code License   & GNU General Public License v3.0 \\
\hline
C5 & Code versioning system used & git \\
\hline
C6 & Software code languages, tools, and services used & R programming language \\
\hline
C7 & Compilation requirements, operating environments \& dependencies & \\
\hline
C8 & If available Link to developer documentation/manual & $https://mayur1009.github.io/cleanTS/$ \\
\hline
C9 & Support email for questions & $mayur.k.shende@gmail.com$, $neerajdhanraj@cae.au.dk$, $afeijoo@uvigo.gal$\\
\hline
\end{tabular}
\caption{Code metadata (mandatory)}
\label{codeMeta} 
\end{table}

% \section*{Current executable software version}
% \label{}

% Ancillary data table required for sub version of the executable software: (x.1, x.2 etc.) kindly replace examples in right column with the correct information about your executables, and leave the left column as it is.

% \begin{table}[!h]
% \begin{tabular}{|l|p{6.5cm}|p{6.5cm}|}
% \hline
% \textbf{Nr.} & \textbf{(Executable) software metadata description} & \textbf{Please fill in this column} \\
% \hline
% S1 & Current software version & For example 1.1, 2.4 etc. \\
% \hline
% S2 & Permanent link to executables of this version  & For example: $https://github.com/combogenomics/$ $DuctApe/releases/tag/DuctApe-0.16.4$ \\
% \hline
% S3  & Permanent link to Reproducible Capsule & \\
% \hline
% S4 & Legal Software License & List one of the approved licenses \\
% \hline
% S5 & Computing platforms/Operating Systems & For example Android, BSD, iOS, Linux, OS X, Microsoft Windows, Unix-like , IBM z/OS, distributed/web based etc. \\
% \hline
% S6 & Installation requirements \& dependencies & \\
% \hline
% S7 & If available, link to user manual - if formally published include a reference to the publication in the reference list & For example: $http://mozart.github.io/documentation/$ \\
% \hline
% S8 & Support email for questions & \\
% \hline
% \end{tabular}
% \caption{Software metadata (optional)}
% \label{} 
% \end{table}

\bibliographystyle{elsarticle-num} 
\bibliography{references}

\end{document}